\documentclass{article}

\PassOptionsToPackage{numbers, compress}{natbib}

\usepackage[preprint]{neurips_2023}

\usepackage[utf8]{inputenc} 
\usepackage[T1]{fontenc}    
\usepackage{hyperref}       
\usepackage{url}            
\usepackage{booktabs}       
\usepackage{amsfonts}       
\usepackage{nicefrac}       
\usepackage{microtype}      
\usepackage{xcolor}         
\usepackage{graphicx}
\usepackage{amsmath}
\usepackage{subfigure}
\usepackage{algorithmic}
\usepackage[ruled,linesnumbered]{algorithm2e}
\usepackage{wrapfig}
\usepackage{color} 

\newcommand{\tabincell}[2]{\begin{tabular}{@{}#1@{}}#2\end{tabular}}

\title{From Explicit Communication to Tacit Cooperation: \\
A Novel Paradigm for Cooperative MARL}
\author{%
  Dapeng Li, Zhiwei Xu, Bin Zhang, Guoliang Fan\\
  Institute of Automation, Chinese Academy of Sciences\\
  School of Artificial Intelligence, University of Chinese Academy of Sciences\\
  \texttt{\{lidapeng2020,xuzhiwei2019,zhangbin2020,guoliang.fan\}@ia.ac.cn} \\
}

\begin{document}

\maketitle

\begin{abstract}
Centralized training with decentralized execution (CTDE) is a widely-used learning paradigm that has achieved significant success in complex tasks. However, partial observability issues and the absence of effectively shared signals between agents often limit its effectiveness in fostering cooperation. While communication can address this challenge, it simultaneously reduces the algorithm's practicality.
Drawing inspiration from human team cooperative learning, we propose a novel paradigm that facilitates a gradual shift from explicit communication to tacit cooperation. 
In the initial training stage, we promote cooperation by sharing relevant information among agents and concurrently reconstructing this information using each agent's local trajectory. We then combine the explicitly communicated information with the reconstructed information to obtain mixed information. Throughout the training process, we progressively reduce the proportion of explicitly communicated information, facilitating a seamless transition to fully decentralized execution without communication.
Experimental results in various scenarios demonstrate that the performance of our method without communication can approaches or even surpasses that of QMIX and communication-based methods.
\end{abstract}

\section{Introduction}
Cooperative multi-agent reinforcement learning (MARL) has made significant progress in practical applications in recent years, such as traffic light control~\cite{trafficlight1,trafficlight2}, autonomous driving~\cite{autonomous_driving}, game playing \cite{dota2drl,starcraft2}, and multi-robot control~\cite{multi-robot1,multi-robot2}. 
To effectively address multi-agent learning problems, various algorithms have emerged. Among these methods, the paradigm of centralized training with decentralized execution (CTDE) has gradually become the most concerned MARL paradigm due to its scalability and ability to handle non-stationary problems. 
The CTDE paradigm serves as a hybrid approach that combines the advantages of both centralized~\cite{centralized_guestrin,central_kok} and decentralized~\cite{indp} learning methods. The fundamental concept of the CTDE paradigm is that agents can access global information in a centralized manner during the training process while operating solely on local observations in a decentralized manner during execution. Based on this approach, many MARL algorithms~\cite{vdn,qmix,qplex,coma,maddpg,mappo,rmaddpg} have demonstrated exceptional performance in complex decision-making tasks.

Despite achieving considerable success, the lack of information sharing among agents can hinder the formation of effective cooperation, particularly in Decentralized Partially Observable Markov Decision Process (Dec-POMDP)~\cite{decpomdp_frans}.
To address this issue, a variety of communication methods~\cite{tarmac,ndq,masia,bicnet,commnet} have been proposed, which facilitate information sharing among agents, such as observations and intentions. These methods aim to enhance agents' understanding of the environment and other agents, ultimately fostering better cooperation patterns. However, a notable drawback of communication methods is that the more information that needs to be transmitted, the higher the communication cost. Moreover, there are often situations where communication is not possible in actual deployment. Therefore, balancing practicality and effectiveness can be seen as a valuable trade-off problem in the MARL domain.

A possible approach that can reconcile this trade-off is each agent uses its local trajectory to reconstruct valuable global information. This approach enables practicality in decentralized systems and better collaboration in communication to be realized simultaneously. However, directly reconstructing global information from local observations is not feasible, and utilizing this inaccurate reconstructed information for decision-making may lead to training instability due to reconstructed bias.


\begin{figure*}[ht]
    \centering
    \includegraphics[width=4.5 in]{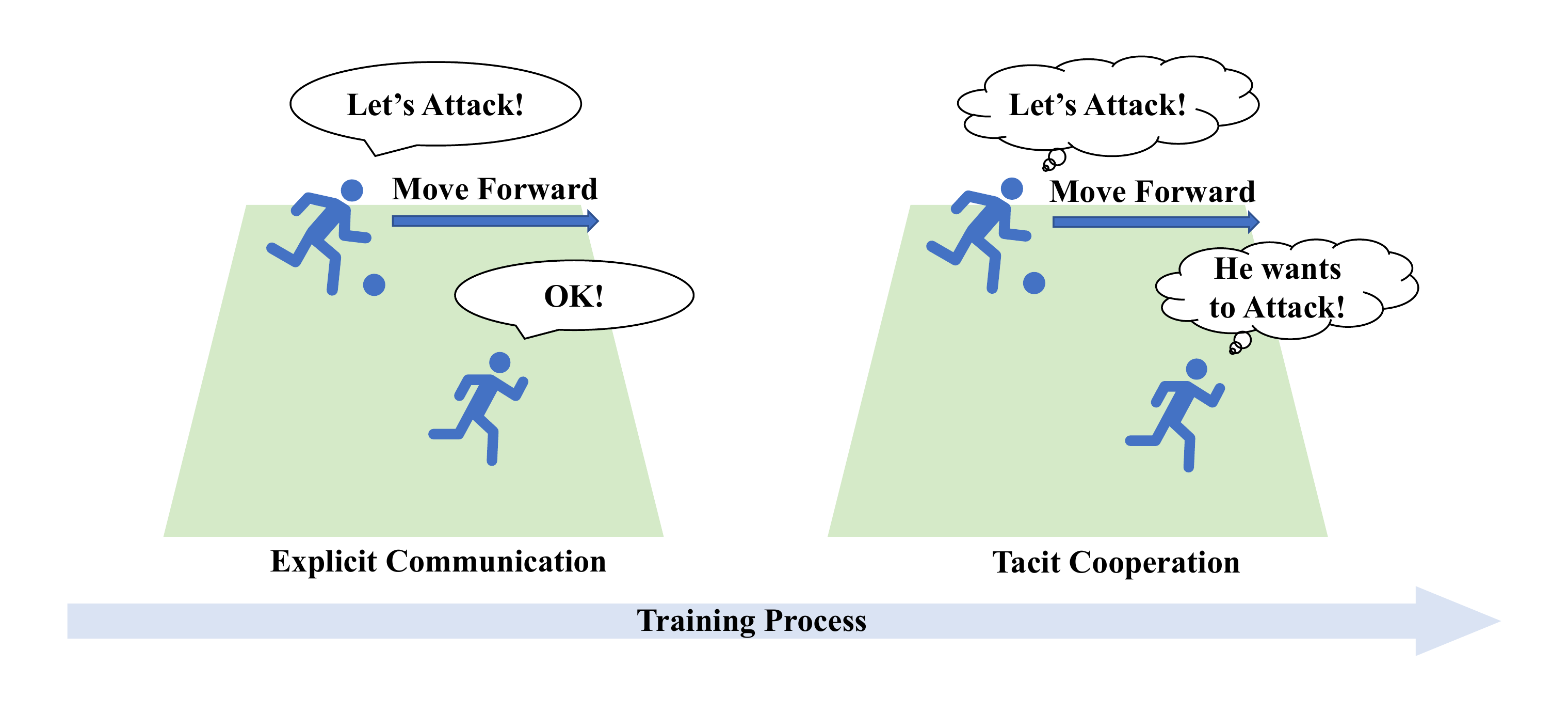}
    \caption{The process of human team cooperative learning.}
    \label{fig:football}
\end{figure*}

In human teamwork (e.g., football and basketball games), a team in the initial stages of training requires communication to foster cooperation, as depicted in the left of Figure~\ref{fig:football}. As team members become more familiar and cohesive, the need for communication gradually reduces, as depicted in the right of Figure~\ref{fig:football}. For example, professional teams can achieve tacit cooperation by only observing the movements of their teammates during real competition.
Drawing inspiration from human teamwork, in this paper, we propose a novel paradigm that can transition from explicit communication to \textbf{TA}cit \textbf{CO}operation~(TACO). This paradigm enables agents to share information via an attention mechanism during the initial training stage while simultaneously reconstructing this information using local observations. We obtain mixed information by weighting and summing the reconstructed information and true global information.
As training progresses, the accuracy of reconstructed information continues to improve. Consequently, we can reduce reliance on actual communication by gradually decreasing its ratio in mixed information, ultimately achieving the ability to infer teammate intentions without actual communication.

We evaluated the performance of TACO on three challenge environments, including StarCraft Multi-Agent Challenge (SMAC)~\cite{smac2}, SMACv2~\cite{smacv2}, and Google Research Football (GRF)~\cite{grf}. The experimental results show that TACO can bring significant improvements to existing CTDE methods and achieve performance comparable to or even superior to communication methods without explicit communication.
Additionally, the visualization results shows that the difference between the reconstructed information and the true global information is relatively small, so it can play a role in promoting cooperation.





\section{Related Work}
MARL has attracted considerable attention and made significant progress in recent years.
To address scalability and non-stationary issues, CTDE flexibly combines the advantages of centralized learning~\cite{centralized_guestrin,central_kok,central_traffic} and independent learning~\cite{indp,stable_replay,dec_marl}, using global information for training and only local observations for execution. Value decomposition is a popular branch of CTDE, with VDN~\cite{vdn} and QMIX~\cite{qmix} being two representative works. VDN avoids the lazy agent problem by decomposing the joint action value function into individual action value functions for each agent. QMIX goes a step further by representing this decomposition process as a monotonic function, leading to improved performance. In a similar vein, MADDPG~\cite{maddpg}, COMA~\cite{coma}, FOP\cite{fop}, MAPPO~\cite{mappo}, MAAC\cite{maac}, and HAPPO~\cite{happo} achieve cooperation through centralized critics. However, the lack of signal sharing between agents during execution significantly limits the performance of CTDE in complex scenarios.

Communication methods address the aforementioned issues by exchanging information among agents. A popular communication approach involves sharing local observations or observation embeddings~\cite{commnet,ic3net,i2c,tarmac,ndq,masia,atoc} to facilitate cooperation. CommNet~\cite{commnet}, for instance, employs the average of hidden layer outputs from all agents as a supplement to local observations. This averaging of hidden layer information, however, may result in information loss and disregard the differences among agents. In practical applications, communication costs often need to be considered, leading many existing methods to focus on developing efficient communication protocols that minimize the communication costs. For example, VBC~\cite{vbc} eliminates noise in messages by incorporating regularization to reduce communication overhead, MAIC~\cite{maic} generates customized information through teammate modeling, and NDQ~\cite{ndq} further compresses message information by maximizing mutual information between messages and agent actions and applying regularization techniques. These studies Indicate the importance of minimizing explicit communication costs for the practical implementation of multi-agent reinforcement learning. 

Unfortunately, communication methods inevitably cannot be directly deployed in environments where communication is not unavailable. Therefore, there is a class of methods that attempt to achieve consensus between agents without explicit communication. COLA~\cite{cola} utilizes the idea of contrastive learning to promote consensus among different agents based on their local observations. PBL~\cite{pbl} attempts to construct implicit representations through actions. MACKRL~\cite{common_knowledge} constructs shared knowledge through a hierarchical policy tree. The main disadvantage of these methods is that consensus formation is only implicitly promoted through the gradient signal during training. Without explicit communication between agents to guide consensus, it is difficult to form good consensus in complex scenarios.
In contrast, our approach is more aligned with human teamwork, where effective collaboration and consensus are first achieved through explicit communication and gradually replaced by the reduction of communication reliance, leading to better tacit collaboration.
\section{Preliminaries}

\newcommand{\tup}[1]{G} 
\newcommand{\Obs}[1]{\Omega_{{#1}}} 
\newcommand{\obs}[1]{o_{{#1}}} 
\newcommand{\obsfunc}[1]{O_{{#1}}} 

\newcommand{\Hidden}[1]{H_{{#1}}} 
\newcommand{\hidden}[1]{h_{{#1}}} %
\newcommand{\State}[1]{S_{{#1}}} 
\newcommand{\state}[1]{s_{{#1}}}
\newcommand{\statetrans}[1]{\mathcal{T}_{{#1}}}

\newcommand{\Act}[1]{U_{{#1}}}
\newcommand{\act}[1]{u_{{#1}}}
\newcommand{\enc}[1]{g_{{#1}}}
\newcommand{\Traj}[1]{\tau_{{#1}}}
\newcommand{\traj}[1]{\tau_{{#1}}}

\newcommand{\agent}[0]{i}

\newcommand{\agentn}[0]{n}
\newcommand{\Agentn}[0]{\mathcal{N}}

\newcommand{\att}[1]{v_{{#1}}}
\newcommand{\rec}[1]{\hat{v}_{{#1}}}

\newcommand{\attweight}[1]{w_{{#1}}}

\newcommand{\reward}[1]{r_{{#1}}} 
\newcommand{\policy}[1]{\pi_{{#1}}} 
\newcommand{\stynum}[0]{K}

\subsection{Problem Formulation}
  
The cooperative multi-agent problem can be considered as a Dec-POMDP game~\cite{decpomdp}. It can be formulated as a tuple $\tup{} = \left< \State{}, \Act{}, \Agentn{}, \Obs{}, \statetrans{}, \obsfunc{}, \reward{},\gamma\right>$, where  $\state{} \in \State{}$ is the true global state of the environment. At each discrete time step $t$, each agent $\agent\in\Agentn{}:=\{1,\dots,\agentn{}\}$ will select an action $\act{\agent}\in \Act{\agent}$.
$\statetrans{}(\state{}'|\state{},\boldsymbol{\act{})}:\State{}\times \Act{} \times \State{} \rightarrow P(\State{})$ is the state transition function, where $\boldsymbol{\act{}}=\{\act{1},\dots,\act{\agentn{}}\}\in\boldsymbol{\Act{}}\equiv\Act{}^\agentn{}$ is the joint action. In the partially observable Markov decision process (POMDP), the global state of the environment is not accessible and each agent $\agent$ can only get its individual observation $\obs{\agent}\in \Obs{}$ by the observation function $\obsfunc{}(\state{}, \agent): \State{} \times \mathcal{N} \rightarrow \Obs{}$. $\reward{\agent}(\state{},\act{\agent}):\State{}\times \Act{\agent} \rightarrow \mathbb{R}$ is the reward function for each agent $\agent$. In some scenarios all agent share the same reward function $\reward{}(\state{},\boldsymbol{\act{}}):\State{}\times \boldsymbol{\Act{}} \rightarrow \mathbb{R}$, i.e. team reward. The goal for each agent is to maximize the expected return, and all agents need to achieve this goal through effective cooperation.



\subsection{Cooperative Multi-Agent Reinforcement Learning}
In a multi-agent system, each agent's policy is continually changing. If other agents are viewed solely as components of the environment, then each agent perceives the environment as unstable from its individual perspective. Therefore, single-agent reinforcement learning algorithms cannot be directly applied to multi-agent tasks. Alternatively, centralized methods consider all agents as a single entity, but this method has the disadvantage of the action space increasing exponentially with the number of agents.

The CTDE paradigm has gained popularity as a more viable multi-agent training method, serving as a compromise between the aforementioned methods. In the centralized training phase, agents can access information such as other agents' observations or policies. However, during decentralized execution, agents can only make decisions based on their local observations. This approach offers advantages in reducing environmental instability and is suitable for handling partially observable multi-agent tasks.

The value decomposition method is based on the CTDE paradigm, which combines individual utility functions $Q_i$ of all agents to fit the global action value $Q_{tot}$. Two classic value decomposition baselines are VDN and QMIX, which decompose $Q_{tot}$ under the assumptions of additivity and monotonicity, respectively. Subsequent value decomposition methods~\cite{qatten,qplex,rode} have employed more complex decomposition techniques but continue to follow the assumption of Individual-Global-Max (IGM). IGM asserts that the overall optimality in multi-agent systems aligns with individual optimality and can be expressed as follows:
\begin{equation}
\arg\max_{\boldsymbol{\act{}}} Q_{tot}(\boldsymbol{\traj{}}, \boldsymbol{\act{}})=\left(\arg \max_{\act{1}} Q_1\left(\traj{1}, \act{1}\right) ,\dots, \arg\max_{\act{\agentn}} Q_N\left(\traj{\agentn}, \act{\agentn}\right)\right),
\end{equation}

where $\boldsymbol{\tau}$ represents the historical trajectory of all agents. In this way, agents in the value decomposition method only need to execute greedy policies, which corresponds to optimal global joint actions. 





\begin{figure*}[t]
	\centering
	\subfigure[]{
		\begin{minipage}[h]{0.31\linewidth}
			\centering
			\includegraphics[width=1.8 in]{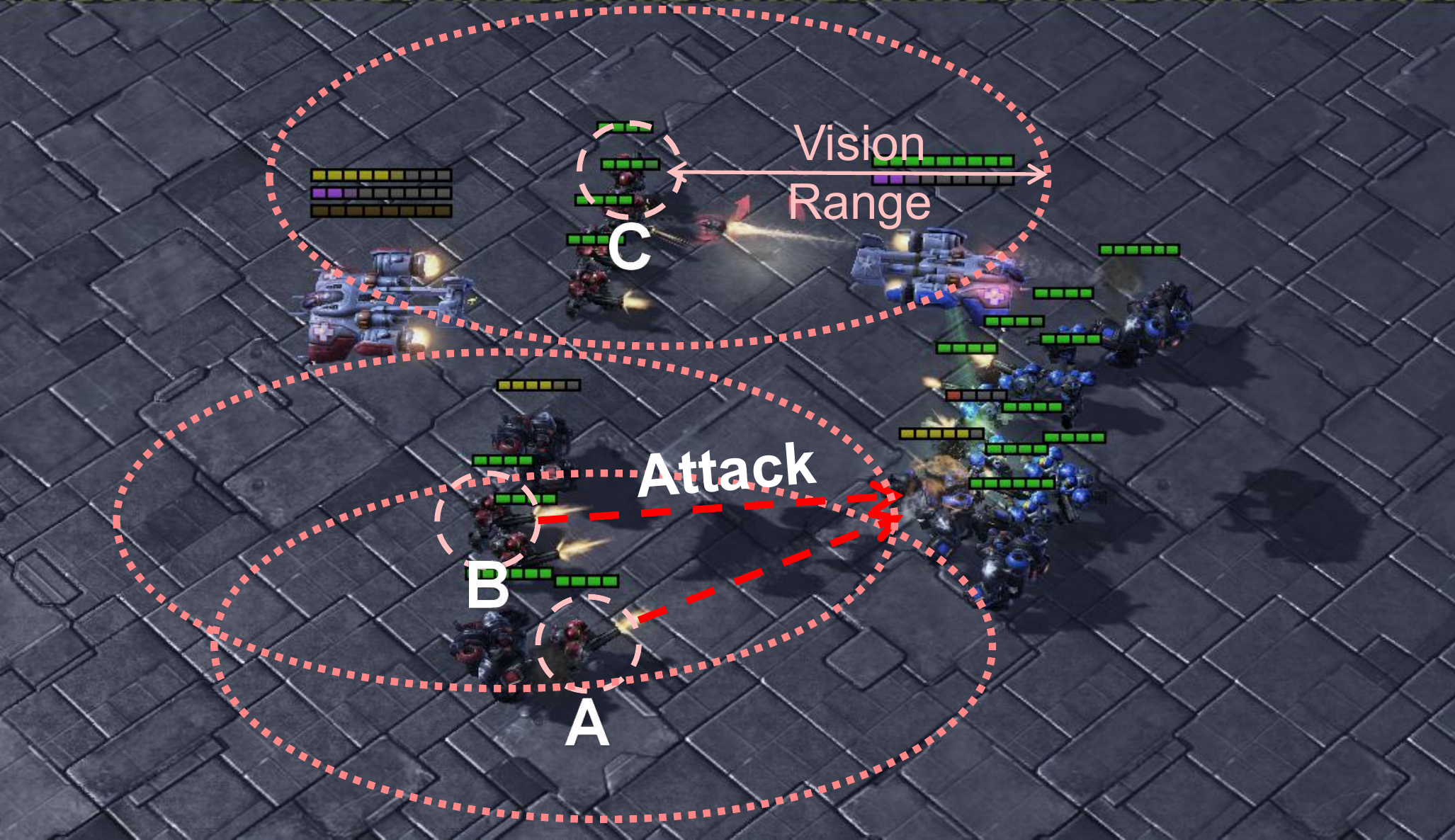}
			\label{fig:MMM2}
		\end{minipage}
	}%
	\subfigure[]{
		\begin{minipage}[h]{0.31\linewidth}
			\centering
			\includegraphics[width=1.8 in]{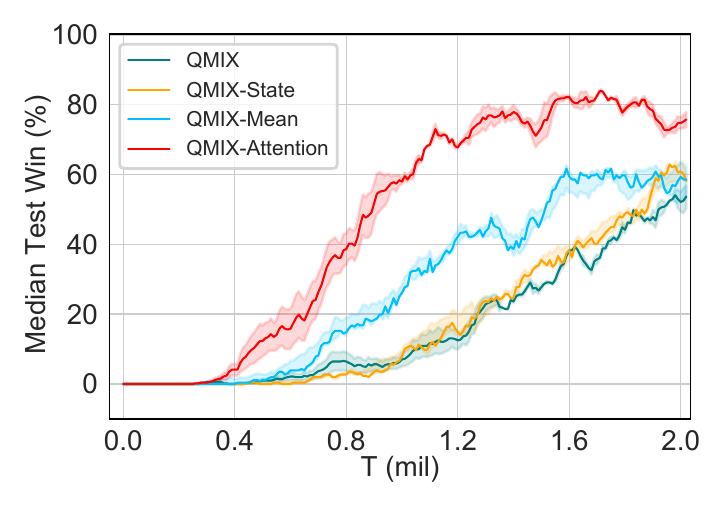}
			\label{fig:motivation_qmix}
		\end{minipage}
	}%
 	\subfigure[]{
		\begin{minipage}[h]{0.31\linewidth}
			\centering
			\includegraphics[width=1.8 in]{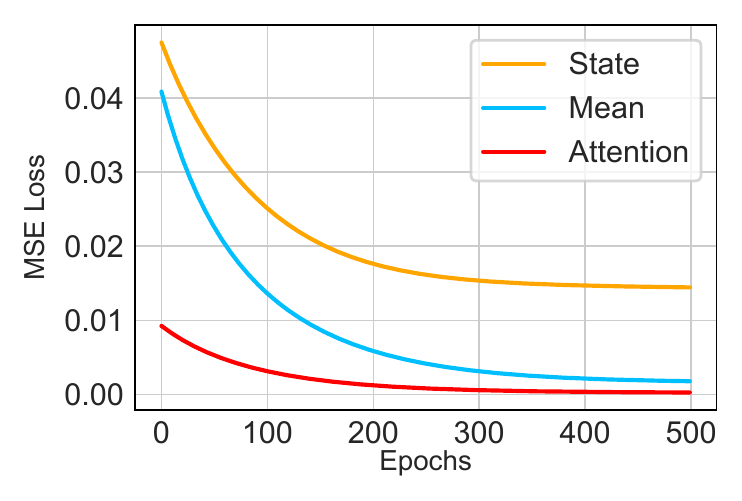}
			\label{fig:mse}
		\end{minipage}
	}%
	\caption{(a) A case study on the \emph{MMM2} scenario. (b) The performance for QMIX with different extra global information. (c) The MSE loss of using local trajectory to reconstruct the different types of global information.}
\label{case study}
\end{figure*}

\section{What Kind of Global Information Do We Need?}
\label{motivation}
To achieve the goal that can balance the trade-off between communication and CTDE, we first need to figure out one question: \textbf{What kind of global information can promote agents' cooperation?}
A naive assumption for this question is that the more information an agent has, the better its decision-making ability. 
However, in fact, receiving too much redundant information in complex environments may hinder the agent from extracting critical information.
We use a motivating example on the \emph{MMM2} of StarCraft to illustrate this viewpoint.
As shown in Figure~\ref{fig:MMM2}, where agents $A$ and $B$ have similar observations due to their proximity. The rational collaborative decision for $A$ and $B$ would be to attack the same enemy simultaneously, highlighting the importance of sharing decision information between them. In contrast, information from another friendly agent $C$, who is not in their field of view, is less important. To verify this motivation, we use QMIX as the baseline and provide agents with different additional global information during execution, including the true global state, the average global feature obtained by mean-weighted hidden states of all friendly agents~(QMIX-Mean), and the attention global feature obtained by weighted hidden states of friendly agents using attention as weights~(QMIX-Attention). Experimental results in Figure~\ref{fig:motivation_qmix} demonstrate that directly providing the global state is not helpful for multi-agent collaboration. Moreover, the improvement obtained by the mean-weighted method is limited, whereas the attention-weighted method significantly enhances the performance of QMIX.

Direct attention communication between agents can be impractical due to high communication costs. However, it is possible to avoid this problem by reconstructing relevant global information based on the local trajectories of agents. This idea is also intuitive: The historical trajectory of agent $A$ includes observations of $B$, and the observations of $A$ and $B$ are very similar too, therefore it is much easier for agent $A$ to reconstruct $B$'s hidden state (or intention) than to reconstruct other irrelevant information (i.e., agent $C$'s intention). Figure~\ref{fig:mse} shows the training Mean Square Error~(MSE) loss for using local observations to reconstruct different global information through a simple Multilayer Perceptron~(MLP). The results show that it is much easier to reconstruct relative attention information than the mean state of all friendly agents and true global states.

Based on the two findings mentioned above (relative information is useful and easy to be reconstructed) and inspired by the progressive cooperation of human beings, we propose the TACO paradigm, which utilizes the extraction of relevant global communication and the reconstruction of relative global communication using local trajectories. This approach involves a gradual transition from communication to tacit cooperation during the training process. In the subsequent sections, we will provide a comprehensive introduction to the TACO paradigm and present the results of our in-depth experimental research.

\section{From Explicit Communication to Tacit Cooperation}
This section aims to provide a comprehensive understanding of TACO, which is a novel multi-agent reinforcement learning paradigm. We will present the framework of TACO and provide a detailed explanation of its implementation to offer a comprehensive overview.

\subsection{The TACO Framework}

\begin{figure*}[t]
    \centering
    \includegraphics[width=5.0 in]{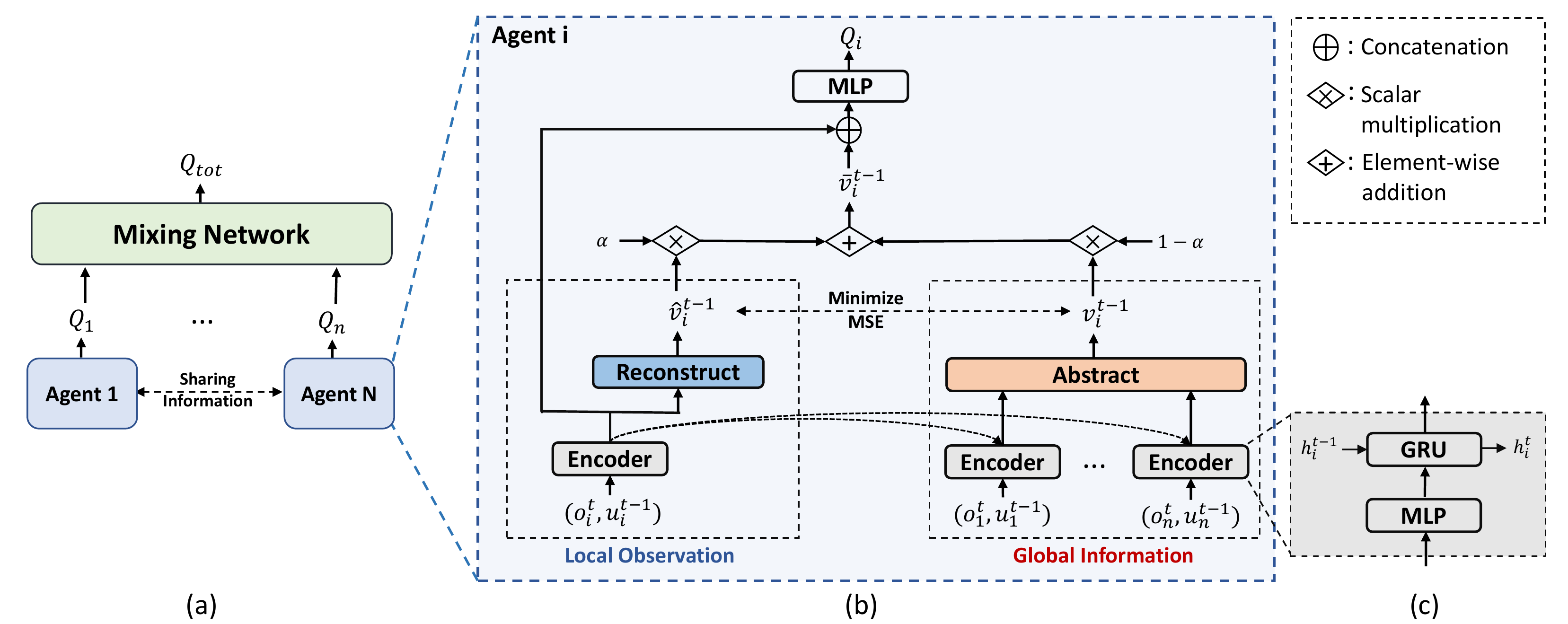}
    \caption{(a) The overall framework of TACO. (b) The detailed agent network. (c) Encoder network structure.}
    \label{fig:structure}
\end{figure*}

To enable a meaningful comparison with other paradigms, the overall structure of TACO adopts the classical value decomposition method, QMIX, which includes both a mixing network and an agent network. 
The core concept of TACO is to gradually reduce the dependence on communication.
As a result, TACO can effectively balance the trade-off between communication and the CTDE method. The entire framework, as depicted in Figure~\ref{fig:structure}(a), consists of a tacit reconstruct module and a communication abstract module, which is detailed in Figure~\ref{fig:structure}(b). Each of these components will be introduced in detail below.

\paragraph{Observation Action Encoder:} The encoder $g$ for each agent $i$ consists of an MLP and a GRU~\cite{gru}, as shown by the gray box in Figure~\ref{fig:structure}(c). It takes the current local observation $\obs{i}^{t}$, previous action $\act{i}^{t}$, and previous hidden state $\hidden{i}^{t-1}$ as input and produces the current hidden state $\hidden{i}^{t} = g(\traj{i}^t, \act{i}^t)$. To enhance learning efficiency, we have set all agents to share the same observation-action encoder. When there is no ambiguity, we omit the superscript of time to simplify notation.

\paragraph{Communication Abstract Module (CAM):}
Given that there are different types of multi-agent relative patterns, such as attention weight between agents, the distance between agents, etc. However, to keep the paper focused, we included only a detailed analysis of attention weights.

The attention weight uses a self-attention mechanism~\cite{attention} to obtain highly relevant global information for each agent. Given the hidden state of agent $i$ and agent $j$, the attention weight $w^{att}_{i,j}$ for agent $i$ to agent $j$ can be computed by using a bilinear mapping and then normalizing it with a softmax function, as shown below:
\begin{equation}
w^{att}_{i,j} = \frac{\textrm{exp}(\hidden{j}^TW^T_kW_q\hidden{i})}{\sum_{j\neq i}\textrm{exp}(\hidden{j}^TW^T_kW_q\hidden{i})},
\end{equation}
In the above equation, $W_k$ and $W_q$ are two learnable linear transformations that represent the "key" function and "query" function, respectively. Furthermore, to extract more diverse global information, we use multiple attention heads and concatenate their outputs together.

Using the relative weight for agent $i$, the global information $v_i$ can be obtained by taking the weighted sum of the hidden state $\hidden{j}$ of all other agents:
\begin{equation}
v_i = \sum_{j\neq i}{w_{i,j}\hidden{j}}.
\end{equation}

\paragraph{Tacit Reconstruct Module (TRM):}
In the TRM, the reconstruct network is responsible for approximating the global attention information $v_i$ based on the local trajectory of each agent. To achieve this, we use a two-layer fully connected network for simplicity. Specifically, the reconstruct network takes the hidden state $\hidden{i}$ of agent $i$ as input and outputs an approximation $\hat{v}_i$ of the global attention information. 

%
\paragraph{From Communication to Tacit Cooperation:}
To ensure a successful transition from communicate to tacit, we obtain the mixed information $\bar{v}_i$ by taking the weighted average of the real attention information $v_i$ and the reconstructed information $\hat{v}_i$:
\begin{equation}
    \bar{v}_i = \alpha \hat{v}_i + (1 - \alpha) v_i.
    \label{average}
\end{equation}
The mixed weight $\alpha$ starts with an initial value $\alpha_{init}$ and using a simple linear increasing schedule, given by $\alpha = \min(\alpha_{init} + t\Delta\alpha,\alpha_{\max})$, which update during each training step. As the $\alpha$ increases, the ratio of the reconstructed information in the mixed information also increases. To ensure the agent can entirely transmit to fully tacit from explicit communication before training is complete, we usually set the $\alpha_{init}=0$, $\alpha_{\max}=1$, and $\Delta{\alpha}=\frac{1}{t_{\max}}$.
The mixed information $\bar{v}_i$ and the hidden state $h_i$ are concatenated to input MLP to obtain $Q_i(\traj{i},\act{i},\bar{v}_i)$. In the scenario where all agents share the same reward, the mixing network decomposes the joint action value function $Q_{tot}$ into the individual action value estimation $Q_i$.

From a communication perspective, the TACO method is advantageous because the model learned does not require actual communication at the end of the training process, which eliminates the need to consider issues related to communication protocols such as bandwidth limitation. This makes TACO more practical compared to other multi-agent reinforcement learning methods that rely on communication.

\subsection{Overall Learning Objective}
We now introduce the learning objectives of TACO, which include two parts: the reinforcement learning part that tries to minimize the TD error, and the mixed information part that attempts to minimize the reconstruct error. 

The reinforcement learning part end-to-end optimizes the same loss function in QMIX~\cite{qmix} which can be formed as follows:
\begin{equation}\mathcal{L}_{\textrm{TD}} = \left(Q_{tot}(\state{},\boldsymbol{\traj{}}, \boldsymbol{\act{}})-y^{tot}\right)^2,
\end{equation}
where $y^{tot}=\reward{}+\gamma \max_{\boldsymbol{\act{}}^{\prime}} Q_{tot}\left(\state{}^{\prime},\boldsymbol{\traj{ }}^{\prime},\boldsymbol{\act{}}^{\prime}\right)$.

To achieve the goal that enforcing the reconstructed information $\hat{v}_i$ to be as close as possible to its corresponding true attention information $v_i$, TACO also includes a mixed information part that minimizes reconstruct loss. The similarity loss between the true global attention information $v_i$ and the reconstructed information $\hat{v}_i$ is measured by using MSE:
\begin{equation}
    \mathcal{L}_{\textrm{Rec}} = \frac{1}{n}\sum^{n}_{i=0}\textrm{MSE}(v_i, \hat{v_i}).
\end{equation}

Both parts are optimized simultaneously during training. Thus, the total loss function can be written as:
\begin{equation}
\mathcal{L}_{tot} = \mathcal{L}_{\textrm{TD}} + \beta\mathcal{L}_{\textrm{Rec}},
\label{total_loss}
\end{equation}

where the $\beta$ is a fixed weighting term. For more difficult scenarios, we can set a higher $\beta$ to increase the proportion of reconstruct loss in the gradient update. It should be noted that the abstract module is updated by both the TD loss and the reconstruct loss gradients in Eq.~\ref{total_loss}. This means that the abstract module generates highly relevant information that is beneficial for agents to collaborate while also making the information easy to be reconstructed.

\section{Experiments}
This section presents experimental results in multiple scenarios that aim to answer the following questions related to TACO:
\textbf{Q1}: Can TACO achieve performance close to that of communication methods in different scenarios?
\textbf{Q2}: Can TACO be applied to different value decomposition baselines to improve their performance?
\textbf{Q3}: Can TACO accurately reconstruct relative attention information from local observations?
\textbf{Q4}: How do the model's hyperparameter settings affect its performance?
We conducted our experiments with StarCraft II~\cite{smac2}, SMACv2~\cite{smacv2}, and Google Research Football~\cite{grf}. All curves are presented with average performance and 25$\sim$75\% deviation over four random seeds, with the solid lines representing the median win rates. 
Additional details about the experiments can be found in Appendix~\ref{Experiment}.

\subsection{Performance on SMAC and SMACv2}
To evaluate our method and baselines, we applied them to the StarCraft II Multi-Agent Challenge benchmark, which includes a series of scenarios representing different levels of challenge, such as easy, hard, and super hard. We select four super hard maps: \emph{3s5z\_vs\_3s6z}, \emph{MMM2}, \emph{27m\_vs\_30m}, \emph{Corridor} and two hard maps: \emph{2c\_vs\_64zg}, \emph{5m\_vs\_6m}. 

To demonstrate the performance of TACO, we used QMIX as the basic framework for TACO and selected two communication methods (NDQ~\cite{ndq} and QMIX-Attention~\cite{rethinking}) and two classic value decomposition methods (QMIX~\cite{qmix} and VDN~\cite{vdn}) as baselines in the main results. Additionally, we compared improved variant of value decomposition methods (QPLEX~\cite{qplex}). 

\begin{figure*}[h]
    \centering
    \includegraphics[width=5.3 in]{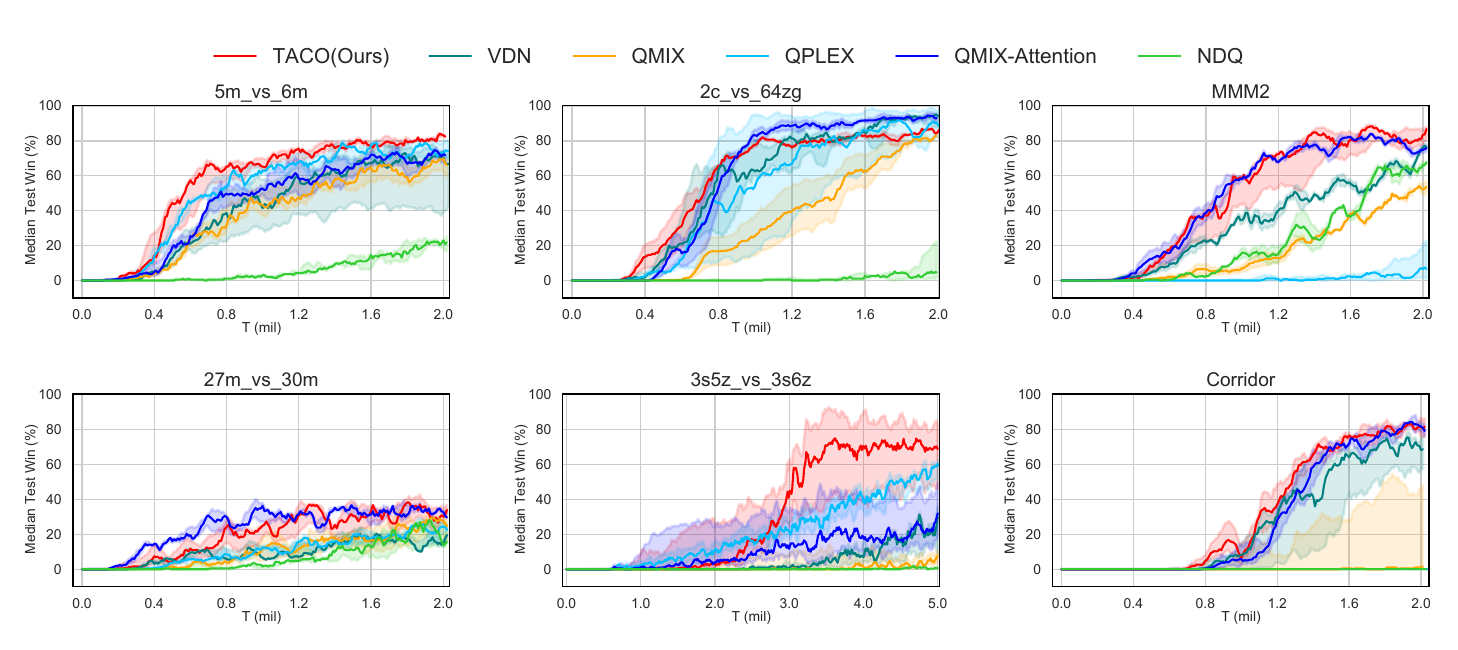}
    \caption{Performance comparison with baselines in different SMAC scenarios.}
    \label{fig:smac_main}
\end{figure*}
In Figure~\ref{fig:smac_main}, it can be observed that there is not much difference between QMIX and QMIX-Attention with full communication in some relatively simple scenarios (\emph{5m\_vs\_6m} and \emph{2c\_vs\_64zg}). However, the NDQ method performs poorly and has low learning efficiency, possibly due to its constraints on message passing and message instability. The performance of TACO is similar to that of QMIX-Attention and even exceeds QMIX-Attention in \emph{5m\_vs\_6m}. 
In the super hard scenarios, the classic CTDE method performs poorly due to a lack of effective communication, whereas QMIX-Attention performs well. However, QMIX-Attention's success is mainly due to its lack of communication restrictions. The TACO method can achieve or even exceed the performance of QMIX-Attention without utilizing actual communication during the end of the training, which significantly enhances its practicality (\textbf{Q1}).  

\begin{figure*}[t]
    \centering
    \includegraphics[width=5.2 in]{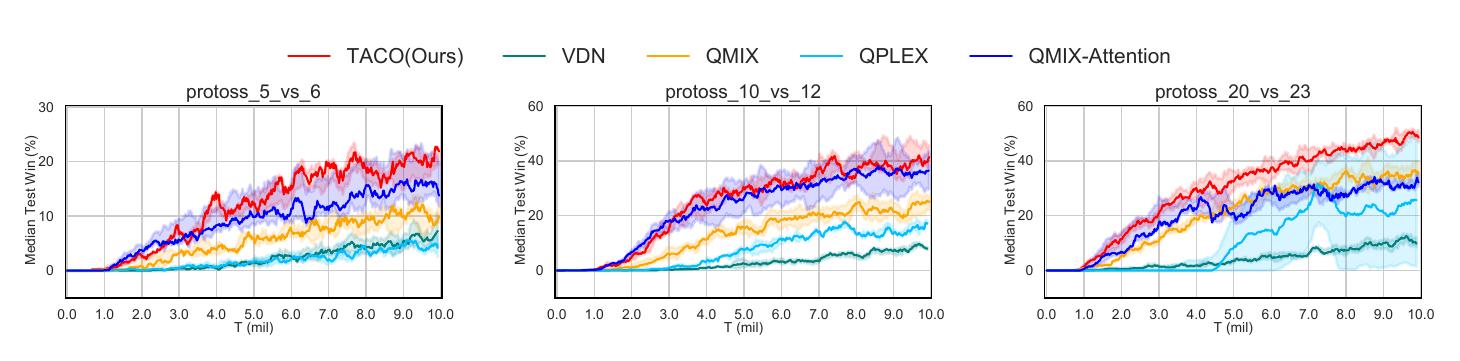}
    \caption{Performance comparison with baselines in different SMACv2 scenarios.}
    \label{fig:smac_mainv2}
\end{figure*}
In SMACv2~\cite{smacv2}, a new improved benchmark, the initial position of each agent is random in each episode, and they have more realistic observations. These changes significantly increase the difficulty of the task and require agents to generalize to previously unseen settings during evaluation. Therefore, we further tested the performance of each algorithm in the SMACv2 environment to demonstrate the generalization and robustness of TACO.
The result in Figure~\ref{fig:smac_mainv2} shows that TACO reaches state-of-the-art performance in SMACv2. Especially, the performances of TACO on three scenarios are even better than QMIX-Attention, which we believe is due to the MSE loss of TACO, which can further promote the extraction of highly correlated global information, especially in complex tasks.

\paragraph{Visualization on SMAC}
\begin{wrapfigure}{r}{0.25\textwidth}
\vspace{-20pt}
\begin{center}
\includegraphics[width=0.25\textwidth]{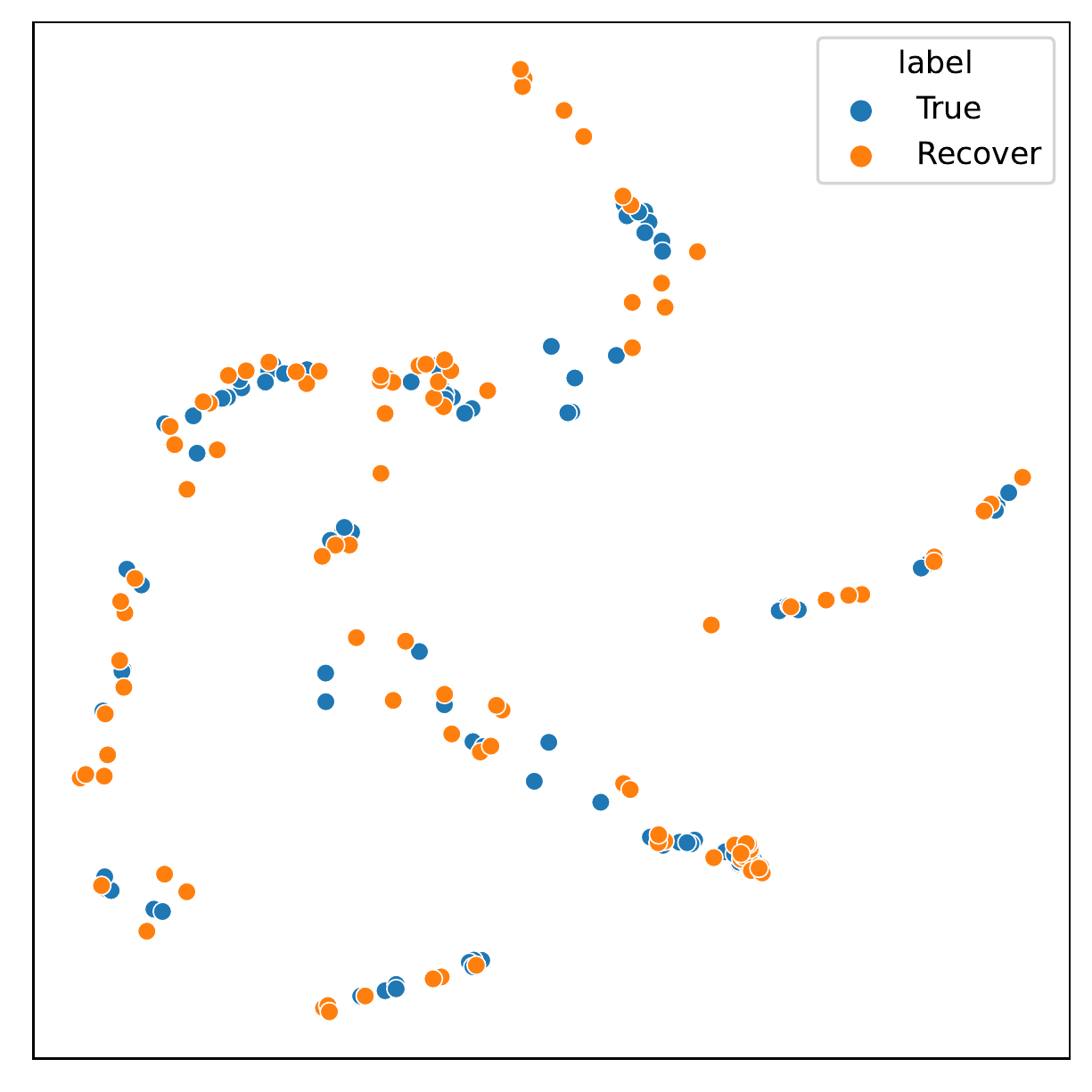}
\end{center}
\vspace{-15pt}
\caption{Visualization.}
\label{visual_tsne}
\end{wrapfigure}

In order to intuitively display the relationship between global features and the reconstructed features, we collected the real global information and global information reconstructed by all agents during the entire episode on \emph{2c\_vs\_64zg}, and used the t-SNE\cite{tsne} to reduce these high-dimensional data into a two-dimensional embedding for visualization.

The t-SNE embedding of true global information (blue) and reconstructed information (orange) are shown in Figure ~\ref{visual_tsne}, it can be seen that although there are some deviations, the reconstructed information can still reflect the distribution of real global information to a certain extent, thereby promoting effective cooperation between agents (\textbf{Q3}).


\subsection{Performance on GRF}
\label{grf}
The GRF environment is a challenging reinforcement learning environment that enables the training of multiple agents to engage in football games. We use VDN, QMIX, VDN-Attention, and QMIX-Attention as baseline models and combined TACO with VDN and QMIX, respectively. To validate the performance of our method, we conducted experiments on three representative official scenarios: \emph{academy\_counterattack\_easy}, \emph{academy\_counterattack\_hard}, and \emph{academy\_run\_pass\_and\_shoot\_with\_keeper}.  

\begin{figure*}[h]
    \centering
    \includegraphics[width=5.2 in]{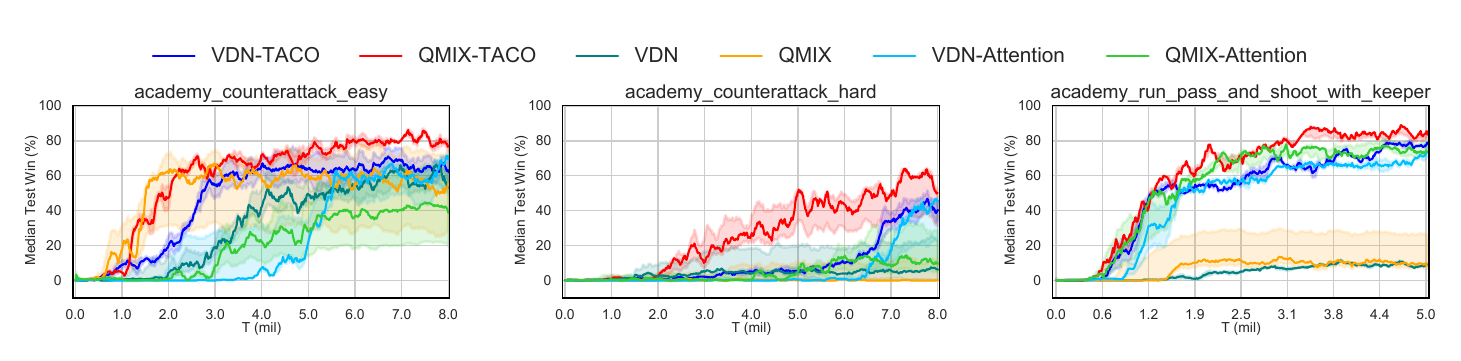}
    \caption{Performance comparison with baselines in different Google Research Football scenarios.}
    \label{fig:grf_main}
\end{figure*}
 As shown in Figure~\ref{fig:grf_main}, baseline models can achieve good results in relatively simple scenarios (i.e. \emph{academy\_counterstack\_easy}), but in more difficult scenarios (i.e. \emph{academy\_counterstack\_hard}), QMIX and VDN perform poorly. QMIX-TACO and VDN-TACO have achieved significant improvements in all three scenarios compared to their baseline models, and are also superior to the QMIX-Attention and VDN-Attention(\textbf{Q2}).


\subsection{Ablation Studies}
In this section, we first tested the impact of several key hyperparameter settings on TACO through ablation experiments (\textbf{Q4}). We selected the \emph{MMM2} scenario of StarCraft and tested the impact of different $\beta$ and attention dimensions on TACO-QMIX.  
\begin{figure*}[ht]
	\centering
	\subfigure[Different beta]{
		\begin{minipage}[h]{0.31\linewidth}
			\centering
			\includegraphics[width=1.8 in]{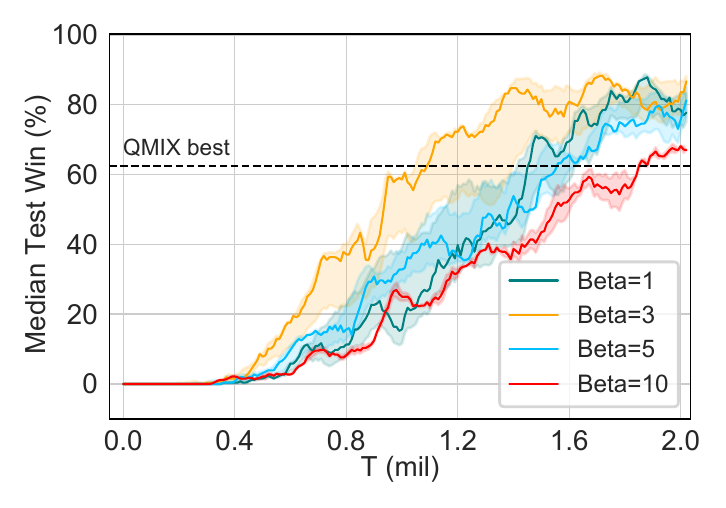}
			\label{fig:MMM2_beta}
		\end{minipage}
	}%
	\subfigure[Different attention dim]{
		\begin{minipage}[h]{0.31\linewidth}
			\centering
			\includegraphics[width=1.8 in]{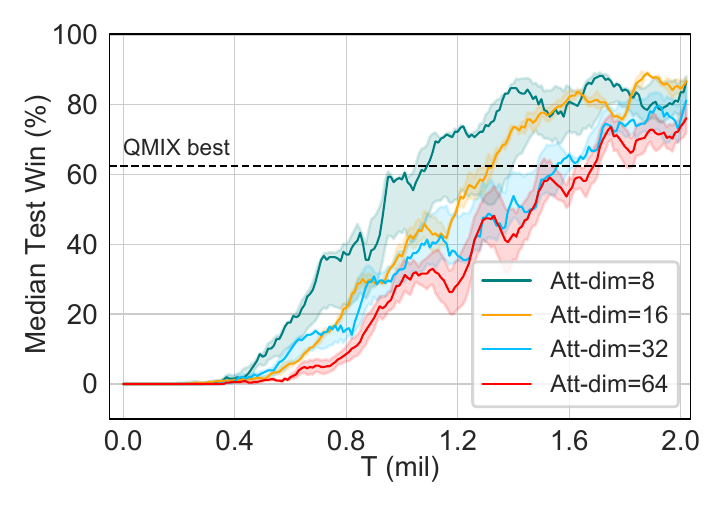}
			\label{fig:MMM2_attdim}
		\end{minipage}
	}%
    \subfigure[The variant of TACO]{
		\begin{minipage}[h]{0.31\linewidth}
			\centering
			\includegraphics[width=1.8 in]{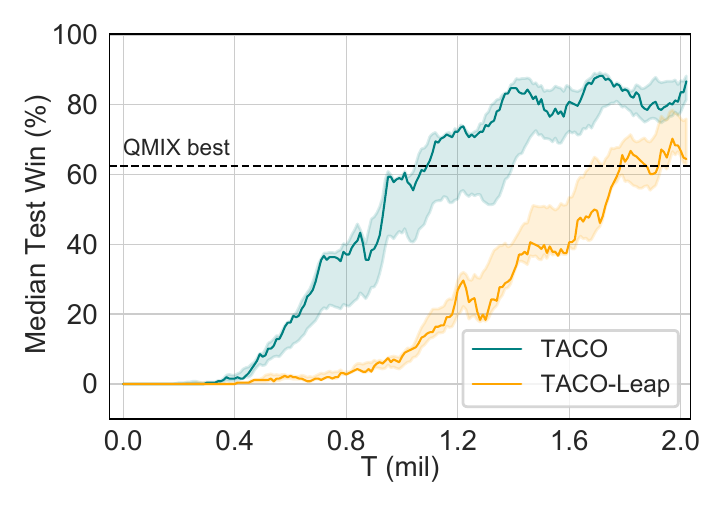}
			\label{fig:MMM2_leap}
		\end{minipage}
	}%
 
	\caption{Ablation studies for TACO on \emph{MMM2} scenario of SMAC.}
\label{Abation study}
\end{figure*}

From Figure~\ref{fig:MMM2_beta}, it can be seen that a too-high $\beta$ value can lead to the model paying too much attention to reconstruct information, thereby hindering the learning of cooperation between agents. From Figure~\ref{fig:MMM2_attdim}, when the dimension of the attention value is lower, better results can be obtained. We believe that this is because lower attention dimensions are more easily reconstructed by local trajectory. The conclusions of the above two parameters in different scenarios are basically the same, but the optimal choice may be different. The detailed hyperparameters for each scenario can be found in Appendix~\ref{hyperparameters}.

Next, we will demonstrate the necessity of a progressive transition process through experiments. To this end, we propose a variant of TACO called TACO-Leap, which uses full communication during the training phase, that is, $\alpha$ is always equal to zero. From Figure~\ref{fig:MMM2_leap}, the process of progressive learning provides a guidance effect, which can promote the learning efficiency of the TACO method and achieve better performance.

\section{Closing Remarks}


In this paper, we propose a simple and effective multi-agent collaboration training paradigm called TACO. This approach allows agents to explicitly share relevant information during the early stages of training to guide effective cooperation, and gradually replace real information with reconstructed information through local trajectories, ultimately achieving efficient cooperation under fully decentralized execution. Experimental results in multiple scenarios show that the TACO method can achieve close or even better performance than the same baseline using communication or global information without sharing information.

We believe that this work is beneficial for the multi-agent collaboration community, as existing communication methods often have practical limitations due to communication bandwidth and other issues, while mainstream CTDE training cannot achieve effective collaboration due to the lack of shared information. Our method opens up a new approach that combines the advantages of the above methods, which is both practical and effective.

It is worth noting that our method may fail when there is totally unrelated between the agents' trajectories or observations, as the effective reconstruction of shared information requires a certain degree of similarity between the historical trajectories of the agents. We will further investigate and attempt to mitigate this limitation in future research.


\medskip

{
\small
\bibliography{ref}

\begin{thebibliography}{47}
\providecommand{\natexlab}[1]{#1}
\providecommand{\url}[1]{\texttt{#1}}
\expandafter\ifx\csname urlstyle\endcsname\relax
  \providecommand{\doi}[1]{doi: #1}\else
  \providecommand{\doi}{doi: \begingroup \urlstyle{rm}\Url}\fi

\bibitem[Arel et~al.(2010)Arel, Liu, Urbanik, and Kohls]{trafficlight1}
Itamar Arel, Cong Liu, Tom Urbanik, and Airton~G Kohls.
\newblock Reinforcement learning-based multi-agent system for network traffic
  signal control.
\newblock \emph{IET Intelligent Transport Systems}, 4\penalty0 (2):\penalty0
  128--135, 2010.

\bibitem[Berner et~al.(2019)Berner, Brockman, Chan, Cheung, Debiak, Dennison,
  Farhi, Fischer, Hashme, Hesse, J{\'{o}}zefowicz, Gray, Olsson, Pachocki,
  Petrov, de~Oliveira~Pinto, Raiman, Salimans, Schlatter, Schneider, Sidor,
  Sutskever, Tang, Wolski, and Zhang]{dota2drl}
Christopher Berner, Greg Brockman, Brooke Chan, Vicki Cheung, Przemyslaw
  Debiak, Christy Dennison, David Farhi, Quirin Fischer, Shariq Hashme,
  Christopher Hesse, Rafal J{\'{o}}zefowicz, Scott Gray, Catherine Olsson,
  Jakub Pachocki, Michael Petrov, Henrique~Pond{\'{e}} de~Oliveira~Pinto,
  Jonathan Raiman, Tim Salimans, Jeremy Schlatter, Jonas Schneider, Szymon
  Sidor, Ilya Sutskever, Jie Tang, Filip Wolski, and Susan Zhang.
\newblock Dota 2 with large scale deep reinforcement learning.
\newblock \emph{CoRR}, abs/1912.06680, 2019.

\bibitem[Bernstein et~al.(2002)Bernstein, Givan, Immerman, and
  Zilberstein]{decpomdp}
Daniel~S. Bernstein, Robert Givan, Neil Immerman, and Shlomo Zilberstein.
\newblock The complexity of decentralized control of markov decision processes.
\newblock \emph{Math. Oper. Res.}, 27\penalty0 (4):\penalty0 819--840, 2002.
\newblock \doi{10.1287/moor.27.4.819.297}.

\bibitem[Cho et~al.(2014)Cho, van Merrienboer, G{\"{u}}l{\c{c}}ehre, Bahdanau,
  Bougares, Schwenk, and Bengio]{gru}
Kyunghyun Cho, Bart van Merrienboer, {\c{C}}aglar G{\"{u}}l{\c{c}}ehre, Dzmitry
  Bahdanau, Fethi Bougares, Holger Schwenk, and Yoshua Bengio.
\newblock Learning phrase representations using {RNN} encoder-decoder for
  statistical machine translation.
\newblock In Alessandro Moschitti, Bo~Pang, and Walter Daelemans, editors,
  \emph{Proceedings of the 2014 Conference on Empirical Methods in Natural
  Language Processing, {EMNLP} 2014, October 25-29, 2014, Doha, Qatar, {A}
  meeting of SIGDAT, a Special Interest Group of the {ACL}}, pages 1724--1734.
  {ACL}, 2014.
\newblock \doi{10.3115/v1/d14-1179}.
\newblock URL \url{https://doi.org/10.3115/v1/d14-1179}.

\bibitem[Das et~al.(2019)Das, Gervet, Romoff, Batra, Parikh, Rabbat, and
  Pineau]{tarmac}
Abhishek Das, Th{\'{e}}ophile Gervet, Joshua Romoff, Dhruv Batra, Devi Parikh,
  Mike Rabbat, and Joelle Pineau.
\newblock Tarmac: Targeted multi-agent communication.
\newblock In Kamalika Chaudhuri and Ruslan Salakhutdinov, editors,
  \emph{Proceedings of the 36th International Conference on Machine Learning,
  {ICML} 2019, 9-15 June 2019, Long Beach, California, {USA}}, volume~97 of
  \emph{Proceedings of Machine Learning Research}, pages 1538--1546. {PMLR},
  2019.

\bibitem[de~Witt et~al.(2019)de~Witt, Foerster, Farquhar, Torr, Boehmer, and
  Whiteson]{common_knowledge}
Christian~Schr{\"{o}}der de~Witt, Jakob~N. Foerster, Gregory Farquhar, Philip
  H.~S. Torr, Wendelin Boehmer, and Shimon Whiteson.
\newblock Multi-agent common knowledge reinforcement learning.
\newblock In Hanna~M. Wallach, Hugo Larochelle, Alina Beygelzimer, Florence
  d'Alch{\'{e}}{-}Buc, Emily~B. Fox, and Roman Garnett, editors, \emph{Advances
  in Neural Information Processing Systems 32: Annual Conference on Neural
  Information Processing Systems 2019, NeurIPS 2019, December 8-14, 2019,
  Vancouver, BC, Canada}, pages 9924--9935, 2019.
\newblock URL
  \url{https://proceedings.neurips.cc/paper/2019/hash/f968fdc88852a4a3a27a81fe3f57bfc5-Abstract.html}.

\bibitem[Devin et~al.(2017)Devin, Gupta, Darrell, Abbeel, and
  Levine]{multi-robot2}
Coline Devin, Abhishek Gupta, Trevor Darrell, Pieter Abbeel, and Sergey Levine.
\newblock Learning modular neural network policies for multi-task and
  multi-robot transfer.
\newblock In \emph{2017 {IEEE} International Conference on Robotics and
  Automation, {ICRA} 2017, Singapore, Singapore, May 29 - June 3, 2017}, pages
  2169--2176. {IEEE}, 2017.
\newblock \doi{10.1109/ICRA.2017.7989250}.
\newblock URL \url{https://doi.org/10.1109/ICRA.2017.7989250}.

\bibitem[Ding et~al.(2020)Ding, Huang, and Lu]{i2c}
Ziluo Ding, Tiejun Huang, and Zongqing Lu.
\newblock Learning individually inferred communication for multi-agent
  cooperation.
\newblock \emph{Advances in Neural Information Processing Systems},
  33:\penalty0 22069--22079, 2020.

\bibitem[Ellis et~al.(2022)Ellis, Moalla, Samvelyan, Sun, Mahajan, Foerster,
  and Whiteson]{smacv2}
Benjamin Ellis, Skander Moalla, Mikayel Samvelyan, Mingfei Sun, Anuj Mahajan,
  Jakob~N. Foerster, and Shimon Whiteson.
\newblock Smacv2: An improved benchmark for cooperative multi-agent
  reinforcement learning.
\newblock \emph{CoRR}, abs/2212.07489, 2022.
\newblock \doi{10.48550/arXiv.2212.07489}.
\newblock URL \url{https://doi.org/10.48550/arXiv.2212.07489}.

\bibitem[Foerster et~al.(2017)Foerster, Nardelli, Farquhar, Afouras, Torr,
  Kohli, and Whiteson]{stable_replay}
Jakob~N. Foerster, Nantas Nardelli, Gregory Farquhar, Triantafyllos Afouras,
  Philip H.~S. Torr, Pushmeet Kohli, and Shimon Whiteson.
\newblock Stabilising experience replay for deep multi-agent reinforcement
  learning.
\newblock In Doina Precup and Yee~Whye Teh, editors, \emph{Proceedings of the
  34th International Conference on Machine Learning, {ICML} 2017, Sydney, NSW,
  Australia, 6-11 August 2017}, volume~70 of \emph{Proceedings of Machine
  Learning Research}, pages 1146--1155. {PMLR}, 2017.
\newblock URL \url{http://proceedings.mlr.press/v70/foerster17b.html}.

\bibitem[Foerster et~al.(2018)Foerster, Farquhar, Afouras, Nardelli, and
  Whiteson]{coma}
Jakob~N. Foerster, Gregory Farquhar, Triantafyllos Afouras, Nantas Nardelli,
  and Shimon Whiteson.
\newblock Counterfactual multi-agent policy gradients.
\newblock In Sheila~A. McIlraith and Kilian~Q. Weinberger, editors,
  \emph{Proceedings of the Thirty-Second {AAAI} Conference on Artificial
  Intelligence, (AAAI-18), the 30th innovative Applications of Artificial
  Intelligence (IAAI-18), and the 8th {AAAI} Symposium on Educational Advances
  in Artificial Intelligence (EAAI-18), New Orleans, Louisiana, USA, February
  2-7, 2018}, pages 2974--2982. {AAAI} Press, 2018.

\bibitem[Guan et~al.(2022)Guan, Chen, Yuan, Wang, Yin, Zhang, and Yu]{masia}
Cong Guan, Feng Chen, Lei Yuan, Chenghe Wang, Hao Yin, Zongzhang Zhang, and
  Yang Yu.
\newblock Efficient multi-agent communication via self-supervised information
  aggregation.
\newblock In Alice~H. Oh, Alekh Agarwal, Danielle Belgrave, and Kyunghyun Cho,
  editors, \emph{Advances in Neural Information Processing Systems}, 2022.
\newblock URL \url{https://openreview.net/forum?id=n4wnZAdBavx}.

\bibitem[Guestrin et~al.(2002)Guestrin, Lagoudakis, and
  Parr]{centralized_guestrin}
Carlos Guestrin, Michail~G. Lagoudakis, and Ronald Parr.
\newblock Coordinated reinforcement learning.
\newblock In Claude Sammut and Achim~G. Hoffmann, editors, \emph{Machine
  Learning, Proceedings of the Nineteenth International Conference {(ICML}
  2002), University of New South Wales, Sydney, Australia, July 8-12, 2002},
  pages 227--234. Morgan Kaufmann, 2002.

\bibitem[Hu et~al.(2021)Hu, Jiang, Harding, Wu, and wei Liao]{rethinking}
Jian Hu, Siyang Jiang, Seth~Austin Harding, Haibin Wu, and Shih wei Liao.
\newblock Rethinking the implementation tricks and monotonicity constraint in
  cooperative multi-agent reinforcement learning.
\newblock 2021.

\bibitem[Iqbal and Sha(2019)]{maac}
Shariq Iqbal and Fei Sha.
\newblock Actor-attention-critic for multi-agent reinforcement learning.
\newblock In Kamalika Chaudhuri and Ruslan Salakhutdinov, editors,
  \emph{Proceedings of the 36th International Conference on Machine Learning,
  {ICML} 2019, 9-15 June 2019, Long Beach, California, {USA}}, volume~97 of
  \emph{Proceedings of Machine Learning Research}, pages 2961--2970. {PMLR},
  2019.

\bibitem[Jiang and Lu(2018)]{atoc}
Jiechuan Jiang and Zongqing Lu.
\newblock Learning attentional communication for multi-agent cooperation.
\newblock In Samy Bengio, Hanna~M. Wallach, Hugo Larochelle, Kristen Grauman,
  Nicol{\`{o}} Cesa{-}Bianchi, and Roman Garnett, editors, \emph{Advances in
  Neural Information Processing Systems 31: Annual Conference on Neural
  Information Processing Systems 2018, NeurIPS 2018, December 3-8, 2018,
  Montr{\'{e}}al, Canada}, pages 7265--7275, 2018.

\bibitem[Kok and Vlassis(2006)]{central_kok}
Jelle~R. Kok and Nikos Vlassis.
\newblock Collaborative multiagent reinforcement learning by payoff
  propagation.
\newblock \emph{J. Mach. Learn. Res.}, 7:\penalty0 1789--1828, 2006.
\newblock URL \url{http://jmlr.org/papers/v7/kok06a.html}.

\bibitem[Kuba et~al.(2022)Kuba, Chen, Wen, Wen, Sun, Wang, and Yang]{happo}
Jakub~Grudzien Kuba, Ruiqing Chen, Muning Wen, Ying Wen, Fanglei Sun, Jun Wang,
  and Yaodong Yang.
\newblock Trust region policy optimisation in multi-agent reinforcement
  learning.
\newblock In \emph{The Tenth International Conference on Learning
  Representations, {ICLR} 2022, Virtual Event, April 25-29, 2022}.
  OpenReview.net, 2022.
\newblock URL \url{https://openreview.net/forum?id=EcGGFkNTxdJ}.

\bibitem[Kurach et~al.(2020)Kurach, Raichuk, Stanczyk, Zajac, Bachem, Espeholt,
  Riquelme, Vincent, Michalski, Bousquet, and Gelly]{grf}
Karol Kurach, Anton Raichuk, Piotr Stanczyk, Michal Zajac, Olivier Bachem,
  Lasse Espeholt, Carlos Riquelme, Damien Vincent, Marcin Michalski, Olivier
  Bousquet, and Sylvain Gelly.
\newblock Google research football: {A} novel reinforcement learning
  environment.
\newblock In \emph{The Thirty-Fourth {AAAI} Conference on Artificial
  Intelligence, {AAAI} 2020, The Thirty-Second Innovative Applications of
  Artificial Intelligence Conference, {IAAI} 2020, The Tenth {AAAI} Symposium
  on Educational Advances in Artificial Intelligence, {EAAI} 2020, New York,
  NY, USA, February 7-12, 2020}, pages 4501--4510. {AAAI} Press, 2020.
\newblock URL \url{https://ojs.aaai.org/index.php/AAAI/article/view/5878}.

\bibitem[Lowe et~al.(2017)Lowe, Wu, Tamar, Harb, Abbeel, and Mordatch]{maddpg}
Ryan Lowe, Yi~Wu, Aviv Tamar, Jean Harb, Pieter Abbeel, and Igor Mordatch.
\newblock Multi-agent actor-critic for mixed cooperative-competitive
  environments.
\newblock In Isabelle Guyon, Ulrike von Luxburg, Samy Bengio, Hanna~M. Wallach,
  Rob Fergus, S.~V.~N. Vishwanathan, and Roman Garnett, editors, \emph{Advances
  in Neural Information Processing Systems 30: Annual Conference on Neural
  Information Processing Systems 2017, December 4-9, 2017, Long Beach, CA,
  {USA}}, pages 6379--6390, 2017.

\bibitem[Matignon et~al.(2012)Matignon, Jeanpierre, and Mouaddib]{multi-robot1}
La{\"{e}}titia Matignon, Laurent Jeanpierre, and Abdel{-}Illah Mouaddib.
\newblock Coordinated multi-robot exploration under communication constraints
  using decentralized markov decision processes.
\newblock In J{\"{o}}rg Hoffmann and Bart Selman, editors, \emph{Proceedings of
  the Twenty-Sixth {AAAI} Conference on Artificial Intelligence, July 22-26,
  2012, Toronto, Ontario, Canada}. {AAAI} Press, 2012.
\newblock URL
  \url{http://www.aaai.org/ocs/index.php/AAAI/AAAI12/paper/view/5038}.

\bibitem[Oliehoek and Amato(2016)]{decpomdp_frans}
Frans~A. Oliehoek and Christopher Amato.
\newblock \emph{A Concise Introduction to Decentralized POMDPs}.
\newblock Springer Briefs in Intelligent Systems. Springer, 2016.
\newblock ISBN 978-3-319-28927-4.
\newblock \doi{10.1007/978-3-319-28929-8}.
\newblock URL \url{https://doi.org/10.1007/978-3-319-28929-8}.

\bibitem[Omidshafiei et~al.(2017)Omidshafiei, Pazis, Amato, How, and
  Vian]{dec_marl}
Shayegan Omidshafiei, Jason Pazis, Christopher Amato, Jonathan~P. How, and John
  Vian.
\newblock Deep decentralized multi-task multi-agent reinforcement learning
  under partial observability.
\newblock In Doina Precup and Yee~Whye Teh, editors, \emph{Proceedings of the
  34th International Conference on Machine Learning, {ICML} 2017, Sydney, NSW,
  Australia, 6-11 August 2017}, volume~70 of \emph{Proceedings of Machine
  Learning Research}, pages 2681--2690. {PMLR}, 2017.
\newblock URL \url{http://proceedings.mlr.press/v70/omidshafiei17a.html}.

\bibitem[Peng et~al.(2017)Peng, Yuan, Wen, Yang, Tang, Long, and Wang]{bicnet}
Peng Peng, Quan Yuan, Ying Wen, Yaodong Yang, Zhenkun Tang, Haitao Long, and
  Jun Wang.
\newblock Multiagent bidirectionally-coordinated nets for learning to play
  starcraft combat games.
\newblock \emph{CoRR}, abs/1703.10069, 2017.
\newblock URL \url{http://arxiv.org/abs/1703.10069}.

\bibitem[Rashid et~al.(2018)Rashid, Samvelyan, de~Witt, Farquhar, Foerster, and
  Whiteson]{qmix}
Tabish Rashid, Mikayel Samvelyan, Christian~Schr{\"{o}}der de~Witt, Gregory
  Farquhar, Jakob~N. Foerster, and Shimon Whiteson.
\newblock {QMIX:} monotonic value function factorisation for deep multi-agent
  reinforcement learning.
\newblock In Jennifer~G. Dy and Andreas Krause, editors, \emph{Proceedings of
  the 35th International Conference on Machine Learning, {ICML} 2018,
  Stockholmsm{\"{a}}ssan, Stockholm, Sweden, July 10-15, 2018}, volume~80 of
  \emph{Proceedings of Machine Learning Research}, pages 4292--4301. {PMLR},
  2018.

\bibitem[Samvelyan et~al.(2019)Samvelyan, Rashid, de~Witt, Farquhar, Nardelli,
  Rudner, Hung, Torr, Foerster, and Whiteson]{smac2}
Mikayel Samvelyan, Tabish Rashid, Christian~Schr{\"{o}}der de~Witt, Gregory
  Farquhar, Nantas Nardelli, Tim G.~J. Rudner, Chia{-}Man Hung, Philip H.~S.
  Torr, Jakob~N. Foerster, and Shimon Whiteson.
\newblock The starcraft multi-agent challenge.
\newblock \emph{CoRR}, abs/1902.04043, 2019.
\newblock URL \url{http://arxiv.org/abs/1902.04043}.

\bibitem[Shalev{-}Shwartz et~al.(2016)Shalev{-}Shwartz, Shammah, and
  Shashua]{autonomous_driving}
Shai Shalev{-}Shwartz, Shaked Shammah, and Amnon Shashua.
\newblock Safe, multi-agent, reinforcement learning for autonomous driving.
\newblock \emph{CoRR}, abs/1610.03295, 2016.

\bibitem[Singh et~al.(2018)Singh, Jain, and Sukhbaatar]{ic3net}
Amanpreet Singh, Tushar Jain, and Sainbayar Sukhbaatar.
\newblock Learning when to communicate at scale in multiagent cooperative and
  competitive tasks.
\newblock \emph{arXiv preprint arXiv:1812.09755}, 2018.

\bibitem[Sukhbaatar et~al.(2016)Sukhbaatar, Szlam, and Fergus]{commnet}
Sainbayar Sukhbaatar, Arthur Szlam, and Rob Fergus.
\newblock Learning multiagent communication with backpropagation.
\newblock In Daniel~D. Lee, Masashi Sugiyama, Ulrike von Luxburg, Isabelle
  Guyon, and Roman Garnett, editors, \emph{Advances in Neural Information
  Processing Systems 29: Annual Conference on Neural Information Processing
  Systems 2016, December 5-10, 2016, Barcelona, Spain}, pages 2244--2252, 2016.
\newblock URL
  \url{https://proceedings.neurips.cc/paper/2016/hash/55b1927fdafef39c48e5b73b5d61ea60-Abstract.html}.

\bibitem[Sunehag et~al.(2017)Sunehag, Lever, Gruslys, Czarnecki, Zambaldi,
  Jaderberg, Lanctot, Sonnerat, Leibo, Tuyls, and Graepel]{vdn}
Peter Sunehag, Guy Lever, Audrunas Gruslys, Wojciech~Marian Czarnecki,
  Vin{\'{\i}}cius~Flores Zambaldi, Max Jaderberg, Marc Lanctot, Nicolas
  Sonnerat, Joel~Z. Leibo, Karl Tuyls, and Thore Graepel.
\newblock Value-decomposition networks for cooperative multi-agent learning.
\newblock \emph{CoRR}, abs/1706.05296, 2017.

\bibitem[Tan(1993)]{indp}
Ming Tan.
\newblock Multi-agent reinforcement learning: Independent versus cooperative
  agents.
\newblock In Paul~E. Utgoff, editor, \emph{Machine Learning, Proceedings of the
  Tenth International Conference, University of Massachusetts, Amherst, MA,
  USA, June 27-29, 1993}, pages 330--337. Morgan Kaufmann, 1993.
\newblock \doi{10.1016/b978-1-55860-307-3.50049-6}.
\newblock URL \url{https://doi.org/10.1016/b978-1-55860-307-3.50049-6}.

\bibitem[Tian et~al.(2020)Tian, Zou, Davies, Warr, Wu, Ammar, and Wang]{pbl}
Zheng Tian, Shihao Zou, Ian Davies, Tim Warr, Lisheng Wu, Haitham~Bou Ammar,
  and Jun Wang.
\newblock Learning to communicate implicitly by actions.
\newblock In \emph{Proceedings of the AAAI Conference on Artificial
  Intelligence}, volume~34, pages 7261--7268, 2020.

\bibitem[Van~der Maaten and Hinton(2008)]{tsne}
Laurens Van~der Maaten and Geoffrey Hinton.
\newblock Visualizing data using t-sne.
\newblock \emph{Journal of machine learning research}, 9\penalty0 (11), 2008.

\bibitem[Van~der Pol and Oliehoek(2016)]{central_traffic}
Elise Van~der Pol and Frans~A Oliehoek.
\newblock Coordinated deep reinforcement learners for traffic light control.
\newblock \emph{Proceedings of learning, inference and control of multi-agent
  systems (at NIPS 2016)}, 8:\penalty0 21--38, 2016.

\bibitem[Vaswani et~al.(2017)Vaswani, Shazeer, Parmar, Uszkoreit, Jones, Gomez,
  Kaiser, and Polosukhin]{attention}
Ashish Vaswani, Noam Shazeer, Niki Parmar, Jakob Uszkoreit, Llion Jones,
  Aidan~N. Gomez, Lukasz Kaiser, and Illia Polosukhin.
\newblock Attention is all you need.
\newblock In Isabelle Guyon, Ulrike von Luxburg, Samy Bengio, Hanna~M. Wallach,
  Rob Fergus, S.~V.~N. Vishwanathan, and Roman Garnett, editors, \emph{Advances
  in Neural Information Processing Systems 30: Annual Conference on Neural
  Information Processing Systems 2017, December 4-9, 2017, Long Beach, CA,
  {USA}}, pages 5998--6008, 2017.
\newblock URL
  \url{https://proceedings.neurips.cc/paper/2017/hash/3f5ee243547dee91fbd053c1c4a845aa-Abstract.html}.

\bibitem[Vinyals et~al.(2019)Vinyals, Babuschkin, Czarnecki, Mathieu, Dudzik,
  Chung, Choi, Powell, Ewalds, Georgiev, Oh, Horgan, Kroiss, Danihelka, Huang,
  Sifre, Cai, Agapiou, Jaderberg, Vezhnevets, Leblond, Pohlen, Dalibard,
  Budden, Sulsky, Molloy, Paine, G{\"{u}}l{\c{c}}ehre, Wang, Pfaff, Wu, Ring,
  Yogatama, W{\"{u}}nsch, McKinney, Smith, Schaul, Lillicrap, Kavukcuoglu,
  Hassabis, Apps, and Silver]{starcraft2}
Oriol Vinyals, Igor Babuschkin, Wojciech~M. Czarnecki, Micha{\"{e}}l Mathieu,
  Andrew Dudzik, Junyoung Chung, David~H. Choi, Richard Powell, Timo Ewalds,
  Petko Georgiev, Junhyuk Oh, Dan Horgan, Manuel Kroiss, Ivo Danihelka, Aja
  Huang, Laurent Sifre, Trevor Cai, John~P. Agapiou, Max Jaderberg,
  Alexander~Sasha Vezhnevets, R{\'{e}}mi Leblond, Tobias Pohlen, Valentin
  Dalibard, David Budden, Yury Sulsky, James Molloy, Tom~Le Paine, {\c{C}}aglar
  G{\"{u}}l{\c{c}}ehre, Ziyu Wang, Tobias Pfaff, Yuhuai Wu, Roman Ring, Dani
  Yogatama, Dario W{\"{u}}nsch, Katrina McKinney, Oliver Smith, Tom Schaul,
  Timothy~P. Lillicrap, Koray Kavukcuoglu, Demis Hassabis, Chris Apps, and
  David Silver.
\newblock Grandmaster level in starcraft {II} using multi-agent reinforcement
  learning.
\newblock \emph{Nat.}, 575\penalty0 (7782):\penalty0 350--354, 2019.
\newblock \doi{10.1038/s41586-019-1724-z}.

\bibitem[Wang et~al.(2021{\natexlab{a}})Wang, Ren, Liu, Yu, and Zhang]{qplex}
Jianhao Wang, Zhizhou Ren, Terry Liu, Yang Yu, and Chongjie Zhang.
\newblock {QPLEX:} duplex dueling multi-agent q-learning.
\newblock In \emph{9th International Conference on Learning Representations,
  {ICLR} 2021, Virtual Event, Austria, May 3-7, 2021}. OpenReview.net,
  2021{\natexlab{a}}.
\newblock URL \url{https://openreview.net/forum?id=Rcmk0xxIQV}.

\bibitem[Wang et~al.(2020{\natexlab{a}})Wang, Everett, and How]{rmaddpg}
Rose~E Wang, Michael Everett, and Jonathan~P How.
\newblock R-maddpg for partially observable environments and limited
  communication.
\newblock \emph{arXiv preprint arXiv:2002.06684}, 2020{\natexlab{a}}.

\bibitem[Wang et~al.(2020{\natexlab{b}})Wang, Wang, Zheng, and Zhang]{ndq}
Tonghan Wang, Jianhao Wang, Chongyi Zheng, and Chongjie Zhang.
\newblock Learning nearly decomposable value functions via communication
  minimization.
\newblock In \emph{8th International Conference on Learning Representations,
  {ICLR} 2020, Addis Ababa, Ethiopia, April 26-30, 2020}. OpenReview.net,
  2020{\natexlab{b}}.
\newblock URL \url{https://openreview.net/forum?id=HJx-3grYDB}.

\bibitem[Wang et~al.(2021{\natexlab{b}})Wang, Gupta, Mahajan, Peng, Whiteson,
  and Zhang]{rode}
Tonghan Wang, Tarun Gupta, Anuj Mahajan, Bei Peng, Shimon Whiteson, and
  Chongjie Zhang.
\newblock {RODE:} learning roles to decompose multi-agent tasks.
\newblock In \emph{9th International Conference on Learning Representations,
  {ICLR} 2021, Virtual Event, Austria, May 3-7, 2021}. OpenReview.net,
  2021{\natexlab{b}}.
\newblock URL \url{https://openreview.net/forum?id=TTUVg6vkNjK}.

\bibitem[Wu et~al.(2017)Wu, Kreidieh, Vinitsky, and Bayen]{trafficlight2}
Cathy Wu, Aboudy Kreidieh, Eugene Vinitsky, and Alexandre~M. Bayen.
\newblock Emergent behaviors in mixed-autonomy traffic.
\newblock In \emph{1st Annual Conference on Robot Learning, CoRL 2017, Mountain
  View, California, USA, November 13-15, 2017, Proceedings}, volume~78 of
  \emph{Proceedings of Machine Learning Research}, pages 398--407. {PMLR},
  2017.
\newblock URL \url{http://proceedings.mlr.press/v78/wu17a.html}.

\bibitem[Xu et~al.(2022)Xu, Zhang, Li, Zhang, Zhou, and Fan]{cola}
Zhiwei Xu, Bin Zhang, Dapeng Li, Zeren Zhang, Guangchong Zhou, and Guoliang
  Fan.
\newblock Consensus learning for cooperative multi-agent reinforcement
  learning.
\newblock \emph{CoRR}, abs/2206.02583, 2022.
\newblock \doi{10.48550/arXiv.2206.02583}.
\newblock URL \url{https://doi.org/10.48550/arXiv.2206.02583}.

\bibitem[Yang et~al.(2020)Yang, Hao, Liao, Shao, Chen, Liu, and Tang]{qatten}
Yaodong Yang, Jianye Hao, Ben Liao, Kun Shao, Guangyong Chen, Wulong Liu, and
  Hongyao Tang.
\newblock Qatten: A general framework for cooperative multiagent reinforcement
  learning.
\newblock \emph{arXiv preprint arXiv:2002.03939}, 2020.

\bibitem[Yu et~al.(2021)Yu, Velu, Vinitsky, Wang, Bayen, and Wu]{mappo}
Chao Yu, Akash Velu, Eugene Vinitsky, Yu~Wang, Alexandre~M. Bayen, and Yi~Wu.
\newblock The surprising effectiveness of {MAPPO} in cooperative, multi-agent
  games.
\newblock \emph{CoRR}, abs/2103.01955, 2021.

\bibitem[Yuan et~al.(2022)Yuan, Wang, Zhang, Wang, Zhang, Yu, and Zhang]{maic}
Lei Yuan, Jianhao Wang, Fuxiang Zhang, Chenghe Wang, Zongzhang Zhang, Yang Yu,
  and Chongjie Zhang.
\newblock Multi-agent incentive communication via decentralized teammate
  modeling.
\newblock In \emph{Thirty-Sixth {AAAI} Conference on Artificial Intelligence,
  {AAAI} 2022, Thirty-Fourth Conference on Innovative Applications of
  Artificial Intelligence, {IAAI} 2022, The Twelveth Symposium on Educational
  Advances in Artificial Intelligence, {EAAI} 2022 Virtual Event, February 22 -
  March 1, 2022}, pages 9466--9474. {AAAI} Press, 2022.
\newblock URL \url{https://ojs.aaai.org/index.php/AAAI/article/view/21179}.

\bibitem[Zhang et~al.(2019)Zhang, Zhang, and Lin]{vbc}
Sai~Qian Zhang, Qi~Zhang, and Jieyu Lin.
\newblock Efficient communication in multi-agent reinforcement learning via
  variance based control.
\newblock \emph{Advances in Neural Information Processing Systems}, 32, 2019.

\bibitem[Zhang et~al.(2021)Zhang, Li, Wang, Xie, and Lu]{fop}
Tianhao Zhang, Yueheng Li, Chen Wang, Guangming Xie, and Zongqing Lu.
\newblock Fop: Factorizing optimal joint policy of maximum-entropy multi-agent
  reinforcement learning.
\newblock In \emph{International Conference on Machine Learning}, pages
  12491--12500. PMLR, 2021.

\end{thebibliography}
\bibliographystyle{plainnat}
}

\newpage
\appendix
\section{Experiment Details}
\label{Experiment}
All experiments in this paper are run on Nvidia GeForce RTX 3090 graphics cards and Intel(R) Xeon(R) Platinum 8280 CPU. Our code is based on PyMARL2~\cite{rethinking}. We fine-tuned the hyperparameter settings of all baselines for a fair comparison. We conducted our experiments using StarCraft II~\cite{smac2}, and Google Research Football~\cite{grf}. 

\begin{figure*}[h]
	\centering
 	\subfigure[StarCraft II]{
		\begin{minipage}[h]{0.47\linewidth}
			\centering
			\includegraphics[width=2.3 in]{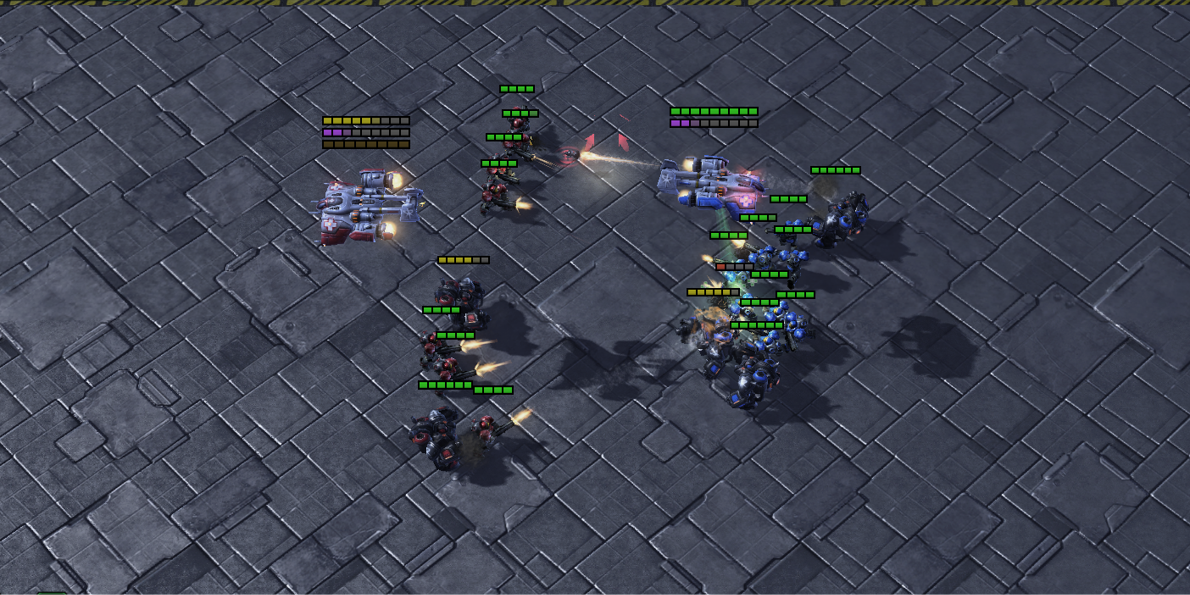}
			\label{fig:visual_smac}
		\end{minipage}
	}%
	\subfigure[Google Research Football]{
		\begin{minipage}[h]{0.47\linewidth}
			\centering
			\includegraphics[width=2.3 in]{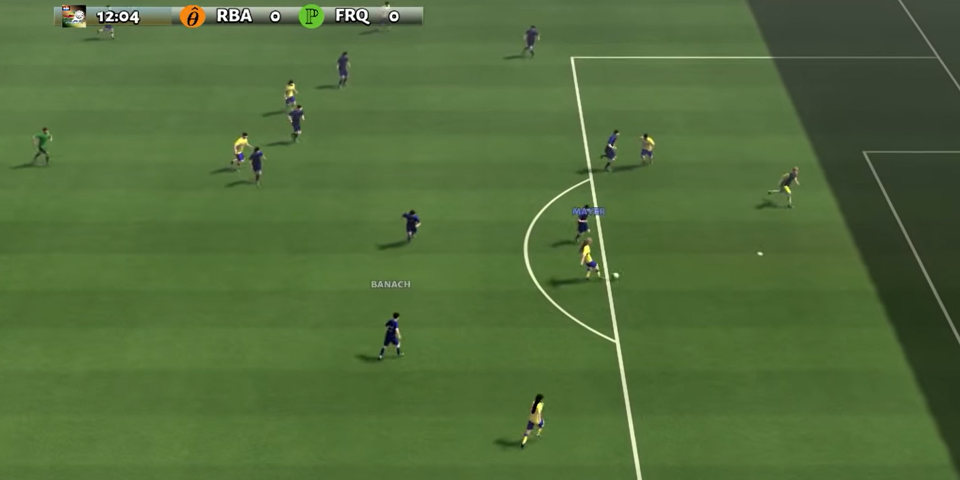}
			\label{fig:visual_grf}
		\end{minipage}
	}%
	\caption{Visualization of experimental environments} 
\label{visual_envs}
\end{figure*}

\subsection{SMAC}
In SMAC, each unit is controlled by an agent and each agent has its vision range. At each time step, the agent can observe all friendly and enemy unit information within its vision range, including distance, relative position, health, shield amount, and unit type. In addition, the agent can obtain a candidate set of currently executable actions. During training, we can utilize the global status, including information about all units on the map. The goal of all friendly agents is to defeat enemy agents in combat, and the reward function is related to the health value of the enemy agent. Table~\ref{table:scenario}provides a detailed description of each SMAC scenario.\footnote{Detail information about SMAC can be found at https://github.com/oxwhirl/smac.}

\begin{table*}[htbp]
\centering
\label{table:scenario}
\caption{SMAC maps in different scenarios.}
\begin{tabular}{lcccc}
\hline
Name&Ally Units&Enemy Units&Type&Difficulty\\
\hline
& & & &  \\[-6pt]
2c\_vs\_64zg&2 Colossi&64 Zerglings&\tabincell{c}{Homogeneous, Asymmetric,\\Large Action space}&Hard \\
\hline
& & & &  \\[-6pt]
5m\_vs\_6m &5 Marines&6 Marines&\tabincell{c}{Homogeneous, Asymmetric,\\Massive Agents}&Hard \\
\hline
& & & &  \\[-6pt]
27m\_vs\_30m&27 Marines&30 Marines&\tabincell{c}{Homogeneous, Asymmetric}&Super Hard \\
\hline
& & & &  \\[-6pt]
MMM2&\tabincell{c}{1 Medivac\\2 Marauders\\7 Marines}&\tabincell{c}{1 Medivac\\3 Marauder\\8 Marines}&\tabincell{c}{Heterogeneous, Asymmetric,\\Macro tactics}&Super Hard\\
\hline
& & & &  \\[-6pt]
3s5z\_vs\_3s6z&\tabincell{c}{3 Stalkers\\ 5 Zealots}&\tabincell{c}{3 Stalkers \\6 Zealots}&\tabincell{c}{Heterogeneous, Asymmetric}&Super Hard\\
\hline
& & & &  \\[-6pt]
Corridor&\tabincell{c}{6 Zealots}&\tabincell{c}{24 Zerglings}&\tabincell{c}{Micro-trick: wall off}&Super Hard\\
\hline
\end{tabular}
\end{table*}

\subsection{SMACv2}
In the collaborative problem of multi-agent reinforcement learning, SMAC is a very popular experimental environment. However, after long-term development and improvement, the difficulty of SMAC is no longer sufficient to distinguish the effectiveness of the new algorithm. Therefore, \citeauthor{smacv2} proposed a new version of the benchmark called SMACv2\footnote{The code for SMACv2 can be found at https://github.com/oxwhirl/smacv2}. In SMACv2, scenarios are procedurally generated for each round, so it is required that the agent can generalize to settings that have not been seen before during testing.
The three main improvements in SMACv2 include random team compositions, random start positions, and increasing diversity of unit types by using true unit attack and sight ranges. Considering that random team composition is difficult to form a stable and effective communication protocol, we did not use a random team composition and keep the remaining two settings, i.e the random starting position and the true attack and sight ranges. The detailed setting of the three maps we used can be found in Table~\ref{table:smacv2_map}.

\begin{table*}[htbp]
\centering
\label{table:smacv2_map}
\caption{SMACv2 maps settings.}
\begin{tabular}{lcccc}
\hline
Name&Ally Units&Enemy Units&Type\\
\hline
& & & &  \\[-6pt]
protoss\_5\_vs\_6 & 5 Stalkers&6 Stalkers&\tabincell{c}{Homogeneous, Asymmetric,\\Random Start Positions, \\True Attack and Sight Ranges} \\
\hline
& & & &  \\[-6pt]
protoss\_10\_vs\_12 & 10 Stalkers&12 Stalkers&\tabincell{c}{Homogeneous, Asymmetric,\\Random Start Positions, \\True Attack and Sight Ranges}\\
\hline
& & & &  \\[-6pt]
protoss\_20\_vs\_23 &20 Stalkers&23 Stalkers&\tabincell{c}{Homogeneous, Asymmetric,\\Random Start Positions, \\True Attack and Sight Ranges}\\
\hline
\end{tabular}
\end{table*}

\subsection{Google Research Football}
In Google Football, we need to control multiple agents on the left to organize collaborative attacks, and the goal of the team is to score goals. Goalkeepers and opposing players are controlled by the game's built-in AI. Each agent can choose 19 actions, including moving forward, tackling, shooting, and passing the ball. The observation of each friendly agent includes the movement direction and position of all players and the ball. A goal scored, as well as a player returning to the left half, or reaching the maximum time step, will end the current round. Our team will get +100 when winning the game, otherwise get -1. 
\begin{figure*}[h]
    \centering
    \includegraphics[width=5.2 in]{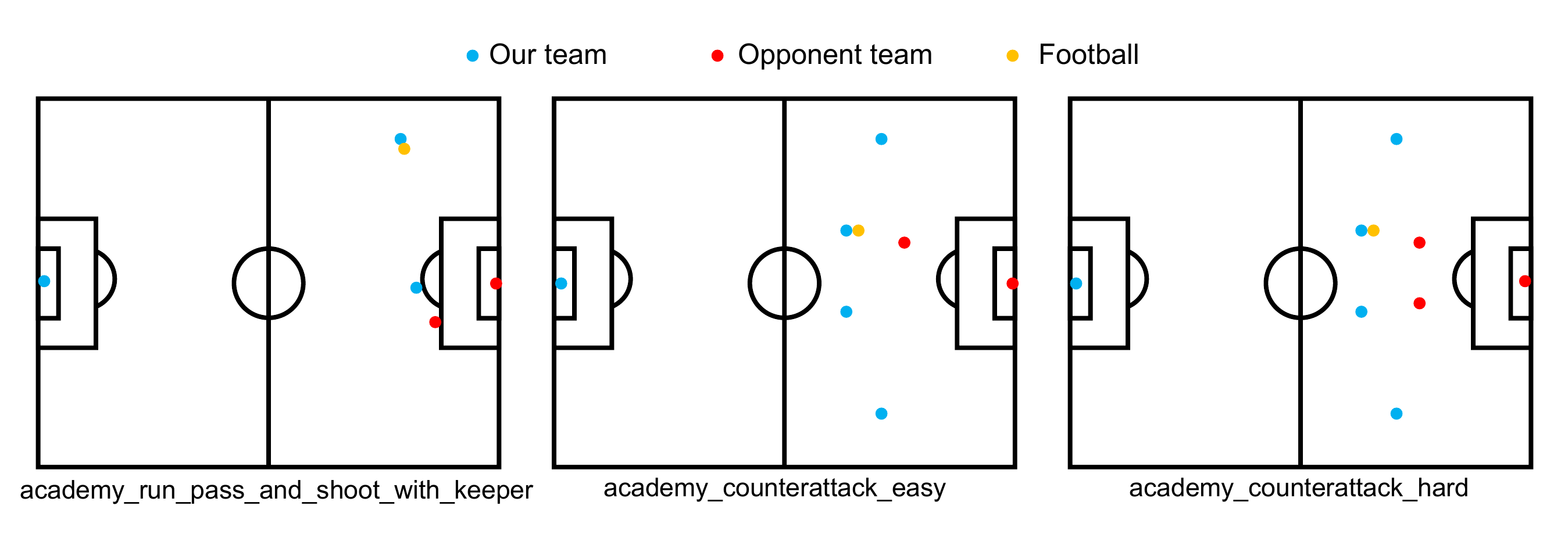}
    \caption{Visualization of the initial position of each agent in three GRF environments considered in our paper.}
    \label{fig:grf_scen}
\end{figure*}

In our paper, we choose three official scenarios in Football Academy\footnote{The official code for Football Academy: https://github.com/google-research/football} including \emph{academy\_run\_pass\_and\_shoot\_with\_keeper}, \emph{academy\_counterattack\_easy}, and \emph{academy\_counterattack\_hard}. 
The initial positions of all players and the ball of all scenarios are shown in Figure~\ref{fig:grf_scen}. In \emph{academy\_counterattack\_easy}, our team has four attacking players, while the enemy has a counter-attack player and a goalkeeper. The \emph{academy\_counterattack\_hard} adds a counter-attack player to the enemy based on \emph{academy\_counterattack\_easy}. In \emph{academy\_run\_pass\_and\_shoot\_with\_keeper}, two of our players try to score, one is on the side with the ball. And the other one is at the center, next to the defender, and facing the opponent keeper.


\subsection{Hyperparameters}
\label{hyperparameters}
We follow the recommendations in \cite{rethinking} to fine-tune the hyperparameters of all baselines. The detailed hyperparameters of TACO and other baselines are shown in Table~\ref{table:hyperparameters}. 
\begin{table}[htbp]
\centering
\setlength{\tabcolsep}{5 mm}{
\begin{tabular}{ll}
\hline
 \textbf{Description}                                        & \textbf{Value} \\ \hline
            Type of optimizer                               & Adam        \\
            Learning rate                                              & 0.001         \\
            How many episodes to update target networks         & 200 \\
            Training epochs of each update                 & 10            \\
            Batch size                                                  & 128             \\
            The capacity of replay buffer (in episodes)                     & 5000           \\
            Starting value for exploration rate annealing   & 1   \\ 
            Ending value for exploration rate annealing   & 20   \\ 
            Discount factor $\gamma$ & 0.99   \\ 
            Dimension of Embedding layer for Mixing network & 32 \\
            Size of hidden state for default GRU cell & 64 \\
            Reduce global norm of gradients & 10 \\
            \hline
            Beta & 3 or 5\\
            Dimension of Attention Embedding layer for Abstract module & 8 or 32\\
            Number of Attention Head & 4 \\
            The maximum value of mixed ratio  $\alpha_{\max}$ & 1 \\
            The initial value of mixed ratio $\alpha_{init}$ & 0 \\
            \hline
            
\end{tabular}}
\caption{Hyperparameters.}
\label{table:hyperparameters}
\end{table}

\section{Pseudo Code of TACO}
The entire algorithm of TACO is presented in Algorithm~\ref{algo:TACO}.
\begin{algorithm}
\label{algo:TACO}
\caption{TACO}
\KwIn{Hyperparameters: $\beta$, $\alpha_{init}$, $\Delta\alpha$.}  
\LinesNumbered
Initialize the replay buffer $D$\\
Initialize the parameters $\theta$ of the agent network and the mixing network\\
Initialize the target parameters $\theta^-$\\

\For{$episode \leftarrow$ 1 \textbf{to} $M$}
{Observe initial state $\boldsymbol{s}^1$ and observation $o^1_a$ for each agent $a$\\
    \For{$t \leftarrow 1$ \textbf{to} $T$}
    {
        \For{$i \leftarrow 1$ \textbf{to} $n$}
        {
        Calculate the mixed information $\bar{v}_i$ by Eq.~\ref{average} \\
        With probability $\epsilon$ select a random action $u_i^t$\\
        Otherwise $u_a^t=\arg\max_{u}Q_a(\tau^t_a, u, \bar{v})$
        }
    Take the joint action $\boldsymbol{u}^t$, and get the next observation $o^{t+1}_i$ and the reward $r^t$\\
    Store the transition $(s^t, \boldsymbol{o}^t, \boldsymbol{u}^t, r^t, s^{t+1}, \boldsymbol{o}^{t+1})$\\
    Update the mixed ratio $\alpha = \min(\alpha{} + \Delta \alpha, \alpha_{\max})$\\
    }
    
Sample a random mini-batch data from $D$\\
Calculate total loss $\mathcal{L}_{tot}$ defined by Eq.~\ref{total_loss} \\
Update $\theta$ by minimizing $\mathcal{L}_{tot}$\\
Update target network parameters $\theta^-=\theta$ periodically}
\end{algorithm}

\end{document}